\documentclass[aps,pra,notitlepage,reprint]{revtex4-1}

\usepackage[colorlinks,linkcolor=red,citecolor=blue,urlcolor=red]{hyperref}
\renewcommand{\t}[1]{\mathrm{#1}}

\usepackage{graphicx}
\usepackage{amsmath,amssymb,amsfonts}

\usepackage{amsmath,amssymb,amsfonts} 
\usepackage{bm}	
\renewcommand{\mathbf}{\bm}

\usepackage{dsfont}	
\renewcommand{\mathbb}{\mathds}	

\usepackage{mathrsfs} 
\usepackage{mathtools} 
\newcommand{\eqdef}{\vcentcolon=\,}

\newcommand{\fref}[1]{Fig.~\ref{#1}}
\renewcommand{\eqref}[1]{Eq.~(\ref{#1})}
\newcommand{\secref}[1]{Sec.~\ref{#1}}

\newcommand{\avg}[1]{\left\langle{#1}\right\rangle}

\newcommand{\re}{\mathrm{Re}\,}
\newcommand{\im}{\mathrm{Im}\,}

\begin{document}

\title{Measurement and control of a mechanical oscillator at its thermal decoherence rate}

\author{D. J. Wilson, V. Sudhir, N. Piro, R. Schilling, A. Ghadimi, and T. J. Kippenberg}

\email{tobias.kippenberg@epfl.ch}

\affiliation{\'{E}cole Polytechnique F\'{e}d\'{e}rale de Lausanne (EPFL), CH-1015 Lausanne, Switzerland}
\date{\today}

\begin{abstract}
In real-time quantum feedback protocols \cite{wiseman_quantum_1994,wiseman2009quantum}, the record of a continuous measurement is used to stabilize a desired quantum state. Recent years have seen highly successful applications in a variety of well-isolated micro-systems, including microwave photons \cite{sayrin_real-time_2011} and superconducting qubits \cite{vijay_stabilizing_2012}.  By contrast, the ability to stabilize the quantum state of a tangibly massive object, such as a nanomechanical oscillator, remains a difficult challenge:  The main obstacle is environmental decoherence, which places stringent requirements on the timescale in which the state must be measured. Here we describe a position sensor that is capable of resolving the zero-point motion of a solid-state, nanomechanical oscillator in the timescale of its thermal decoherence, a critical requirement for preparing its ground state using feedback \cite{doherty_quantum_2012}. The sensor is based on cavity optomechanical coupling \cite{aspelmeyer_cavity_2013}, and realizes a measurement of the oscillator's displacement with an imprecision $40$ dB below that at the standard quantum limit \cite{caves_quantum-mechanical_1980}, 
while maintaining an imprecision-back-action product within a factor of 5 of the Heisenberg uncertainty limit. Using the measurement as an error signal and radiation pressure as an actuator, we demonstrate active feedback cooling (cold-damping \cite{courty_quantum_2001}) of the 4.3 MHz oscillator from a cryogenic bath temperature of  4.4 K to an effective value of 1.1$\pm$0.1 mK, corresponding to a mean phonon number of 5.3$\pm$0.6 (i.e., a ground state probability of 16$\%$). Our results set a new benchmark for the performance of a linear position sensor, and signal the emergence of engineered mechanical oscillators as practical subjects for measurement-based quantum control.
\end{abstract}

\maketitle

{\small
{Feedback control of mechanical oscillators has a long tradition, dating back to steam governors \cite{maxwell_j_clerk_governors_1867}, mechanical clockworks \cite{headrick2002origin} and deflection galvanometers \cite{sirs_galvanometer_1959}.  A basic approach uses a sensor to track the oscillator's position and an actuator to convert the measurement record into a continuous, `real-time' feedback force. Recently the quantum limits of real-time feedback \cite{wiseman_quantum_1994,wiseman2009quantum} have been explored in the context of well-isolated, individual quantum systems, realizing spectacular applications such as generation of microwave Fock states \cite{sayrin_real-time_2011} and persistent Rabi oscillations in a superconducting qubit \cite{vijay_stabilizing_2012}.   In these protocols, the basic paradigm involves a `weak measurement' capable of tracking a quantum state as rapidly as it decoheres due to measurement back-action \cite{hatridge2013quantum}.  For mechanical oscillators, ideal weak position measurements \cite{clerk_introduction_2010} have in fact been available since the advent of the laser, in the context of shot-noise-limited interferometry \cite{caves_quantum-mechanical_1980}.  Only recently, however, with the confluence of low-loss, cryogenic micromechanics and on-chip, integrated photonics \cite{aspelmeyer_cavity_2013}, has it
 been feasible to consider their application to quantum feedback protocols \cite{courty_quantum_2001,szorkovszky2011mechanical}. 
The main challenge is that for a typical, radio-frequency micromechanical oscillator, the thermal environment constitutes an additional, strong decoherence channel.  In order to control a micromechanical oscillator using measurement-based quantum feedback, it is necessary that the measurement be weak (minimally invasive) and yet at the same time efficient enough to resolve the oscillator's quantum state within its thermal decoherence time.  This places stringent demands on the measurement precision. 

Feedback cooling is a well-studied \cite{cohadon_cooling_1999,courty_quantum_2001,mancini_optomechanical_1988,doherty_quantum_2012} control protocol that illustrates both the utility and the challenge of quantum feedback applied to mechanical systems.  In feedback cooling protocols, a mechanical oscillator undergoing thermal Brownian motion is steered towards its ground state by minimizing a measurement of its displacement, $S_x$ (here expressed as a spectral density evaluated at the mechanical frequency, $\Omega_\t{m}$).  The conventional strategy \cite{courty_quantum_2001} is to apply a feedback force which is proportional to the oscillator's velocity, thereby damping the motion until it coincides with the measurement imprecision, $S_x^\t{imp}$.  Ground state cooling (an oscillator phonon occupancy of $n_\t{m}<1$) is possible when the imprecision remains lower than the zero-point fluctuations of the damped oscillator, i.e., if $S_x^{\t{imp}}\lesssim S_x^\t{zp}/n_\t{th}$ (see S.I.)
where $S_x^\t{zp}$ is the oscillator's undamped zero-point displacement and $n_\t{th}$ is the thermal bath occupation.  Practically, this amounts to resolving the undamped thermal noise, $S_x\simeq2n_{\t{th}}S_{x}^{\t{zp}}$, with a signal-to-noise greater than 2$n_{\t{th}}^2$. Equivalently, it corresponds to the ability to resolve the zero-point motion of the oscillator at a characteristic measurement rate \cite{clerk_introduction_2010},
\begin{equation}\label{eq:measurementrate}
\Gamma_{\t{meas}}\equiv \frac{x_\t{zp}^2}{2S_x^\t{imp}}\gtrsim\frac{\Gamma_{\t{th}}}{8},
\end{equation}
where $x_\t{zp}$ is the oscillator's zero-point amplitude, $\Gamma_\t{th}\simeq\Gamma_\t{m}n_\t{th}$ is its thermal decoherence rate and $\Gamma_\t{m}$ is its intrinsic mechanical damping rate (note $S_x^\t{zp}=4x_\t{zp}^2/\Gamma_\t{m}$; see S.I.).
Meeting the requirement set by \eqref{eq:measurementrate} is a daunting technical challenge, owing to the large thermal occupation and small zero-point amplitude of typical micromechanical oscillators. 
\begin{figure*}[t!]
\centering
\includegraphics[width=0.95\linewidth]{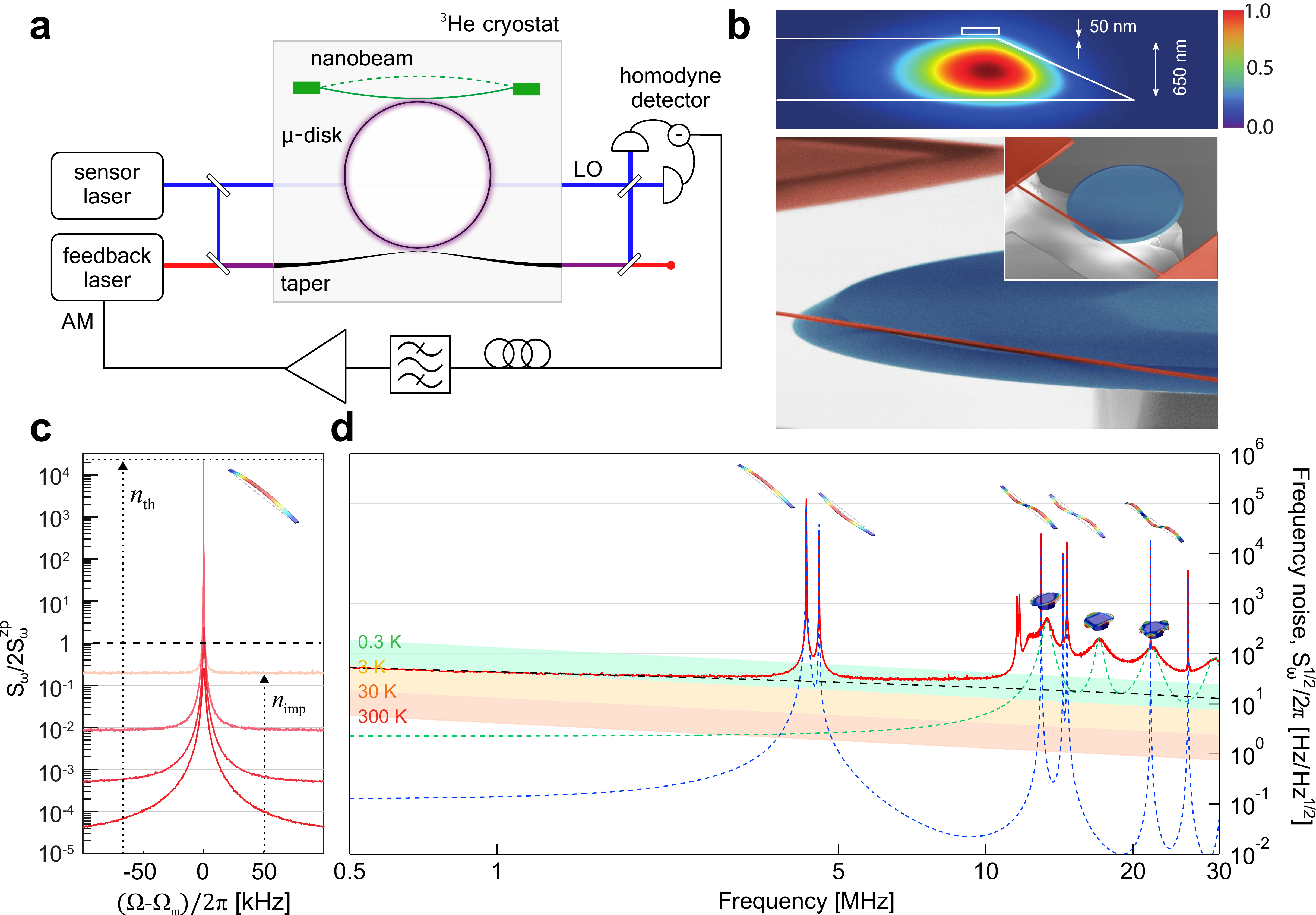}
\caption{\textbf{Measuring and controlling the position of a nanomechanical beam using a near-field optomechanical transducer.} (a) Whispering gallery modes of a SiO$_2$ microdisk are excited using a tapered optical fiber driven by a pair of tunable diode lasers.  Displacement of a Si$_3$N$_4$ nanobeam, sampling the evanescent mode volume of the microdisk, is recorded in the phase of the transmitted sensor field using a balanced homodyne detector.  Radiation pressure feedback is applied by modulating the amplitude of the feedback laser  with an electrically processed (delayed, bandpass-filtered, and amplified) copy of the homodyne photocurrent. (b) Above: Finite element model of the optical mode (field amplitude). Optomechanical coupling is proportional to the field intensity gradient at the position of the beam.  Below: SEM image of the optomechanical system.  (c) Thermomechanical noise spectrum of the fundamental beam mode, measured with varying intracavity photon number.  (d)  Broadband extraneous (shot-noise-subtracted) homodyne signal expressed as apparent cavity frequency noise.  Solid red corresponds to measurement data. Dashed blue, green, and black lines correspond to estimated contributions from nanobeam thermomechanical, microdisk thermomechanical, and microdisk thermorefractive noise, respectively. Colored bands denote the imprecision required for $\Gamma_{\t{meas}}=\Gamma_{\t{th}}$: $S_\omega^{\t{imp}}=g_0^2\hbar Q_{\t{m}}/2k_B T$, assuming $g_0^2\propto 1/\Omega$ and $Q_{\t{m}} = 7.6\cdot 10^5$.}
\label{fig:SetupFigure}
\end{figure*}

\begin{figure*}[t!]
\centering
\includegraphics[width=0.95\linewidth]{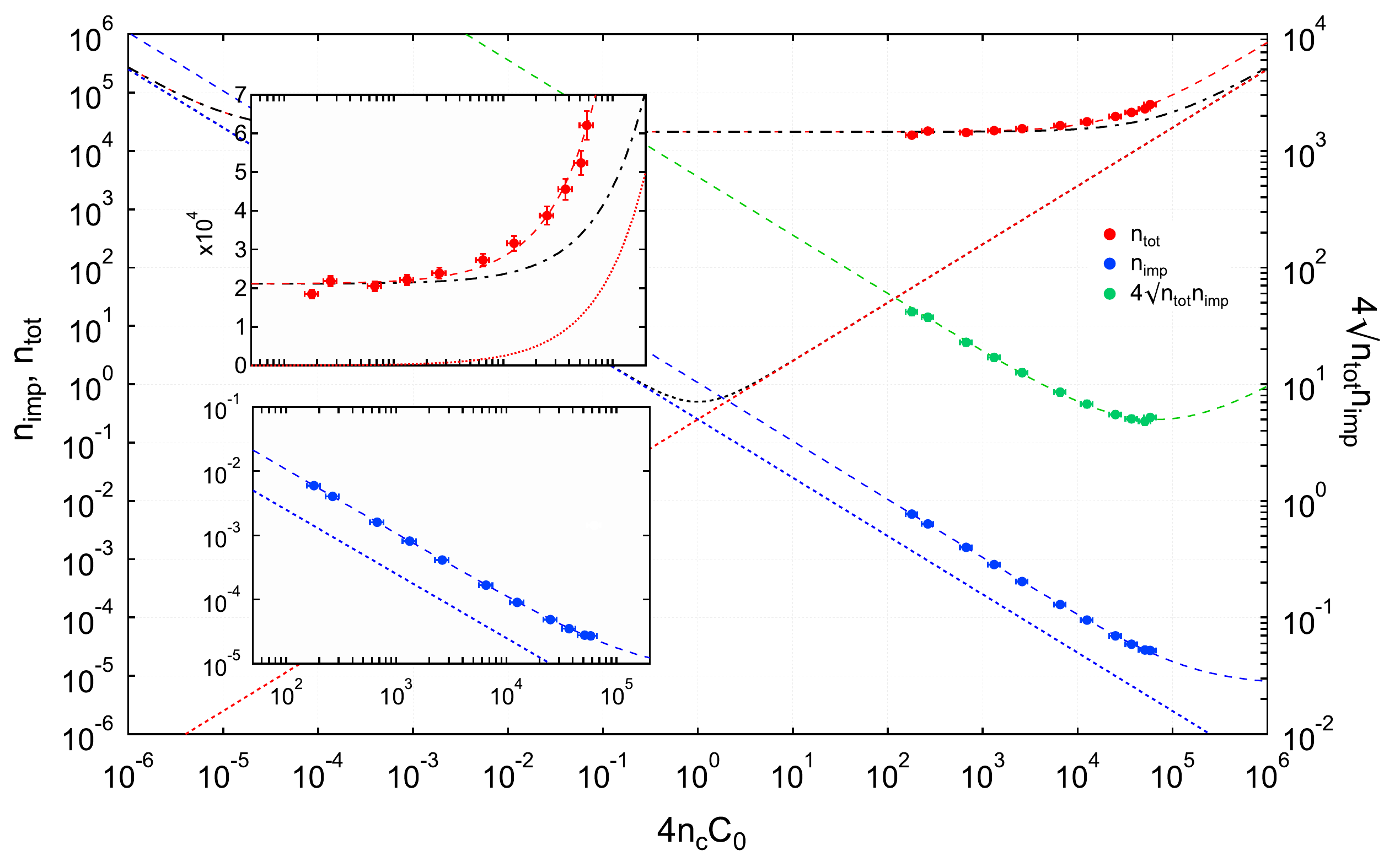}
\caption{\textbf{Measurement imprecision and back-action versus intracavity photon number}. Red, blue, and green points correspond to measurements of total effective bath occupation, $n_\t{tot}=n_{\t{th}}+n_{\t{ba}}$, measurement imprecision referred to an equivalent bath occupation, $n_{\t{\t{imp}}}$, and the apparent imprecision-back-action product, $4\sqrt{n_\t{tot}n_{\t{\t{imp}}}}$, respectively. Short-dashed red, blue, and black lines correspond to models of $n_{\t{ba}}=C_0 n_\t{c}$, $n_\t{imp}=1/16C_0 n_\t{c}$, and their sum, $n_{\t{ba}}+n_{\t{\t{imp}}}$, respectively.  Dash-dotted black line represents the apparent bath occupation $n_\t{tot}+n_\t{imp}$. Long-dashed red and blue lines highlight excursion from their counterparts due to extraneous back-action, $C_0^\t{ex}= 0.56$, extraneous imprecision, $n_{\t{imp}}^\t{ex}= 0.70\cdot10^{-5}$, and imperfect detection efficiency, $\xi= 0.23$, as described in the main text. Green line models the apparent imprecision-back-action product using the Eq. \ref{eq_product}.  Insets highlight the measurement region.} 
\label{fig:imprecision_backaction_figure}
\end{figure*}

An additional, fundamental caveat at once compounds the challenge of feedback cooling and hints at the underlying virtue of quantum feedback:  Heisenberg's uncertainty principle predicts that a weak ($\Gamma_\t{meas}\ll\Omega_\t{m}$) continuous position measurement \cite{clerk_introduction_2010} with an imprecision of $S_{x}^{\t{zp}}/2$ will produce a stochastic `back-action' force that disturbs the position of the oscillator by at least the same amount \cite{caves_quantum-mechanical_1980,clerk_introduction_2010}.
By inference, an imprecision of $n_{\t{\t{imp}}} \equiv S_x^{\t{\t{imp}}}/2S_{x}^\t{zp}$ equivalent bath quanta results in an effective increase of the thermal bath occupation by $n_{\t{ba}}\ge 1/16n_{\t{\t{imp}}}$ (see S.I.).  This penalty would appear to prohibit ground-state cooling, as it entails substantially heating the oscillator to achieve the necessary measurement precision. Remarkably, however, feedback counteracts back-action \cite{wiseman_using_1995}, so that a phonon occupancy of $n_\t{m}\approx 2\sqrt{n_{\t{\t{imp}}}(n_{\t{ba}}+n_{\t{th}})}-1/2< 1$ (see S.I.) can still be achieved \cite{courty_quantum_2001,genes_ground-state_2008,mancini_optomechanical_1988}.  The limiting case of $n_\t{m}\rightarrow0$ is approached when the measurement record is dominated by back-action-induced fluctuations. This occurs when the measurement is maximally efficient \cite{hatridge2013quantum}, i.e., when the measurement rate,  $\Gamma_\t{meas}=\Gamma_\t{m}/16n_\t{imp}$, approaches the effective thermal decoherence rate, $\Gamma_\t{tot}=(n_\t{th}+n_\t{ba})\Gamma_\t{m}\ge\Gamma_\t{meas}$.  To meet this condition for a typical micromechanical oscillator, a linear position sensor must achieve an imprecision far ($\sim n_\t{th}$ times) below the natural scale set by the `standard quantum limit' (SQL) \cite{caves_quantum-mechanical_1980} ($n_\t{imp}=n_\t{ba}=1/4$), while maintaining back-action near the uncertainty limit: $4\sqrt{n_\t{ba}n_\t{imp}}\ge 1$.
 
Integration of micromechanical oscillators with optical and microwave cavities has emerged as a promising pathway to meeting the above requirements. Transduction in such `cavity-optomechanical' systems \cite{aspelmeyer_cavity_2013} arises from a parametric coupling, $G = \partial \omega_{\t{c}}/\partial x$, between the position of the oscillator and the resonance frequency $\omega_{\t{c}}$ of a cavity. For wide-band sensing applications, characterized by a cavity decay rate $\kappa\gg\Omega_\t{m}$, a resonant laser field passing through the cavity acquires a phase shift $2G \delta x/\kappa$; this can be resolved in a conventional homodyne interferometer with a quantum-noise-limited imprecision of 
$S_x^\t{imp}=(8G^2n_c\eta/\kappa)^{-1}$, where $n_\t{c}$ is the mean intracavity photon number and $\eta\in[0,1]$ is the effective photon collection efficiency (see S.I.). The associated quantum-limited measurement rate is given by $\Gamma_{\t{meas}}= 4g_0^2n_\t{c}\eta/\kappa\equiv \Gamma_\t{m}\cdot C_{0} n_\t{c}\eta $, where $g_0\equiv G x_\t{zp}$ is the vacuum optomechanical coupling rate and $C_{0}\equiv 4g_{0}^2/\kappa\Gamma_\t{m}$, the `single-photon cooperativity'  \cite{aspelmeyer_cavity_2013}, characterizes the per-photon measurement rate. To achieve efficient measurements, contemporary cavity-optomechanical systems build on relentless progress in the NEMS/MEMS and photonics communities --- dovetailing fabrication techniques which enable substantial miniaturization of the mechanical resonator and the optical cavity while reinforcing low-loss and strong co-localization.  As a consequence, measurements with an imprecision below that at the SQL \cite{teufel_nanomechanical_2009,anetsberger_measuring_2010,westphal_interferometer_2012}, as well as quantum-back-action (i.e., radiation pressure shot noise) limited measurements \cite{murch_observation_2008,purdy_observation_2013,safavi-naeini_squeezed_2013,suh_mechanically_2014} have recently been realized.  In none of these experiments, however, was $\Gamma_\t{meas}\approx \Gamma_\t{th}$ demonstrated at the detector, owing to a combination of factors including large thermal occupation, extraneous imprecision, optical loss, and dynamic instabilities.

\begin{figure*}
\centering
\includegraphics[width=0.95\linewidth]{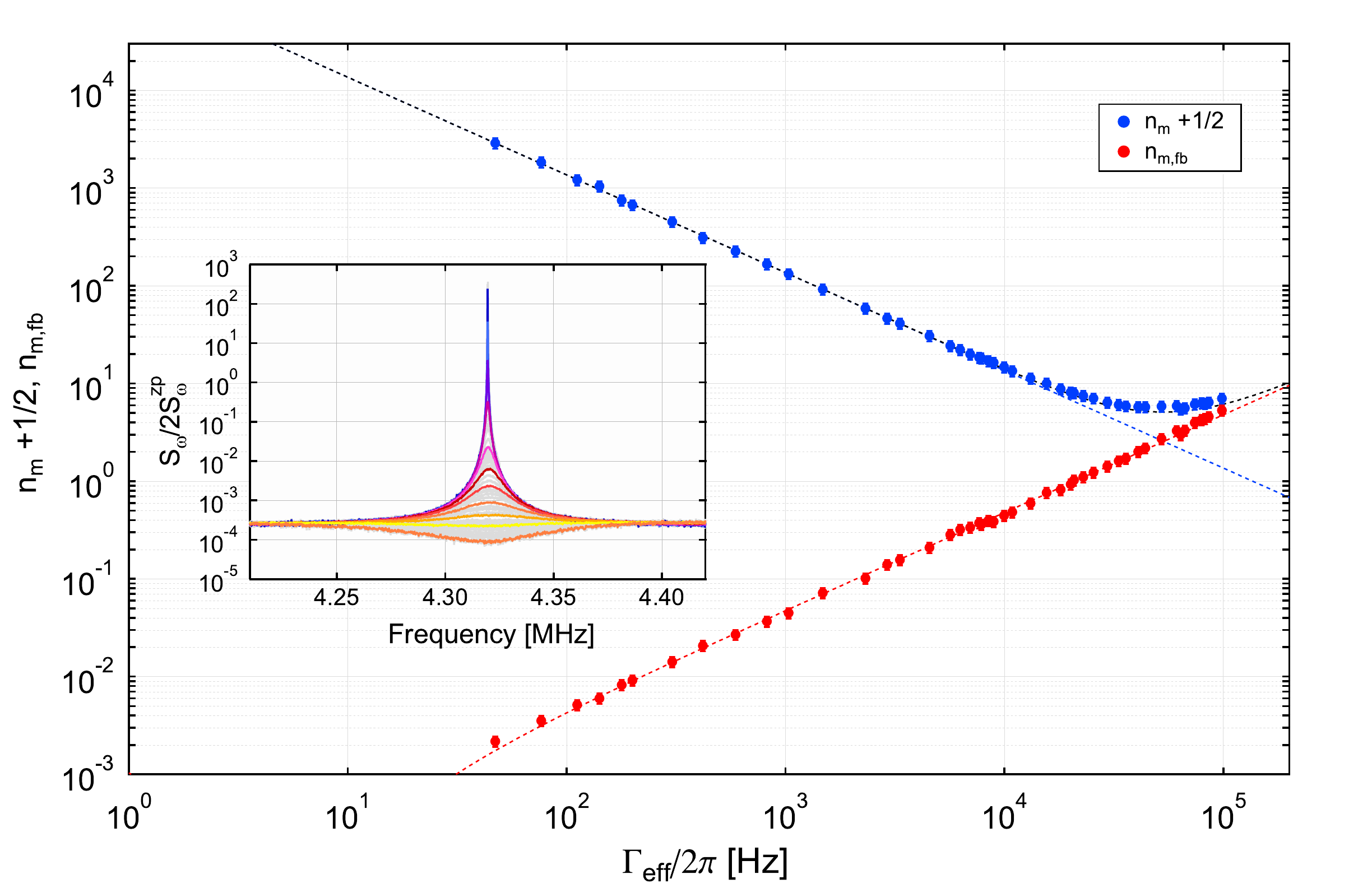}
\caption{\textbf{Radiation pressure feedback cooling to near the ground state}. Blue and red points correspond to measurements of the phonon occupancy of the mechanical mode, $n_\t{m}$ (plus a phonon-equivalent zero-point energy of $1/2$) and its component due to feedback of measurement noise $n_{\t{m,fb}}=n_\t{imp}g_\t{fb}^2/(1+g_\t{fb})$, respectively, as a function of effective damping rate, $\Gamma_{\t{eff}}=(1+g_\t{fb})\Gamma_\t{m}$. Red, blue, and black dashed lines correspond to models of components in Eq. \ref{eq_cooling}: $n_\t{tot}/(1+g_\t{fb})$, $n_\t{m,fb}$, and $n_\t{m}+1/2$, respectively, using experimental parameters $\Gamma_{\t{m}} = 2\pi\cdot 5.7$ Hz, $n_\t{tot}=2.4\cdot 10^5$, and $n_{\t{\t{imp}}} = 2.9\cdot 10^{-4}$, respectively. Inset: in-loop mechanical noise spectra for various feedback gain settings; fits to these spectra were used to infer blue and red points.}
\label{fig:feedbackcoolingfigure}
\end{figure*}

Our system addresses these challenges by exploiting a form of optomechanical coupling --- near-field coupling \cite{anetsberger_near-field_2009} --- that allows for integration of mechanical and optical resonators with widely differing material and geometry.  To achieve high cooperativity, we integrate a mechanical oscillator possessing an exceptionally high $Q$/(mass) ratio and low optical absorption 
--- a high-stress Si$_{3}$N$_{4}$ nanomechanical beam \cite{verbridge_high_2006} --- with an optical cavity possessing a high $Q$/(mode volume) ratio and low optical nonlinearity --- a chemically polished SiO$_2$ microdisk \cite{kippenberg2006demonstration}.  As visualized in Fig. \ref{fig:SetupFigure}b, coupling is achieved by carefully localizing a portion of the beam within the evanescent volume of one of the microdisk's whispering gallery modes.  Both resonators are integrated on a silicon chip \cite{gavartin_hybrid_2012}, allowing for robust cryogenic operation.  

Specifically, we study a system consisting of a 65 $\mu$m $\times$  400 nm $\times$ 70 nm (effective mass $m\approx 2.9$ pg) nanobeam placed $\sim$ 50 nm from the surface of a 30 $\mu$m diameter microdisk.  The microdisk is optically probed using a low-loss ($\approx 6\%$) fiber-taper \cite{spillane_ideality_2003} and light supplied by a tunable diode laser.  Mechanical motion is observed in the phase of the transmitted cavity field using a balanced homodyne interferometer.  We interrogate two optical modes: a `sensor' mode (used for homodyne readout) at $\lambda_{\t{c}}\approx 775$ nm that exhibits an intrinsic photon decay rate of $\kappa_0\approx 2\pi\cdot 0.44$ GHz 
and a `feedback' mode (used for radiation pressure actuation) at $\lambda_{\t{c}}\approx 843$ nm that exhibits a decay rate of $\kappa_0\sim 2\pi\cdot 1$ GHz.  For the mechanical oscillator, we use the $\Omega_{\t{m}}\approx 2\pi\cdot 4.3$ MHz fundamental out-of-plane mode of the nanobeam.  The optomechanial coupling strength between the oscillator and the sensor mode is $g_\t{0}\approx2\pi\cdot20$ kHz, corresponding to a frequency pulling factor of $G\approx 2\pi\cdot 0.70$ GHz/nm for the estimated zero-point amplitude of $x_{\t{zp}}\equiv \sqrt{\hbar/2m\Omega_\t{m}}\approx 29$ fm. Our experiments are conducted in a $^{3}$He buffer gas cryostat at a nominal operating temperature of $T\approx 4.4$ K ($n_{\t{th}}\simeq k_\t{B}T/\hbar\Omega_\t{m}\approx 2.1\cdot 10^4$) and at gas pressures below $10^{-3}$ mbar.  Ringdown measurements here reveal a mechanical damping rate of $\Gamma_{\t{m}}\approx 2\pi\cdot 5.7$ Hz ($Q_{\t{m}}\approx 7.6\cdot 10^{5}$).  Our system is thus able to operate with a near-unity single-photon cooperativity of $C_{0}\approx 0.64$.

For all position sensors, extraneous thermal fluctuations place a fundamental limit on the achievable precision. For cavity-optomechanical sensors, the main sources of extraneous imprecision arise from thermomechanical  \cite{gillespie1993thermal,arcizet_high-sensitivity_2006} and thermorefractive \cite{braginsky2000thermo} fluctuations of the cavity substrate.  These result in excess cavity frequency noise, $S_\omega^\t{imp,ex}$, and limit the measurement rate to
\begin{equation}\label{eq:practicalmeasurementrate}
\Gamma_\t{meas}=\frac{g_{0}^2/2}{S_\omega^\t{imp,shot}+S_\omega^\t{imp,ex}}=\frac{\Gamma_\t{m}/16}{n_\t{imp}^\t{shot}+n_\t{imp}^\t{ex}}
\end{equation}
where $S_\omega^\t{imp,shot}$ is the photocurrent shot noise referred to apparent cavity resonance frequency noise and $n_\t{imp}^\t{shot(ex)}\equiv S_\omega^{\t{imp,shot(ex)}}/2S_\omega^{\t{zp}}$.  Fig. \ref{fig:SetupFigure}d shows the extraneous noise floor of our sensor over a broad range of frequencies surrounding the fundamental beam resonance.  We obtained this spectrum by subtracting shot noise from a measurement made with a large intracavity photon number, $n_\t{c}>10^5$.  (To mitigate thermo-optic and optomechanical instabilities, the measurement was in this case conducted using $\sim10$ mbar of gas pressure at an elevated temperature of 15.7 K.) 
High- and low-Q noise peaks correspond to thermal motion of the nanobeam and the microdisk, respectively. In the vicinity of the fundamental noise peak, we observe an extraneous frequency noise background of  $S^\t{imp,ex}_{\omega}\approx (2\pi\cdot  30 \t{Hz}/\sqrt{\t{Hz}})^2$, corresponding to an extraneous position imprecision of $S_x^{\t{imp,ex}}\approx (4.3\cdot 10^{-17} \t{m}/\sqrt{\t{Hz}})^2$. We identify this noise as a combination of microdisk thermorefractive noise \cite{anetsberger_measuring_2010}, diode laser frequency noise \cite{wieman1991using}, and off-resonant thermal motion of the neighboring in-plane beam mode at 4.6 MHz. Owing to the large zero-point motion of the oscillator, $S_\omega^\t{zp}\equiv4g_0^2/\Gamma_{\t{m}} = (2\pi\cdot 6.7\;\t{ kHz}/\sqrt{\t{Hz}})^2$ ($S_x^{\t{zp}}= (0.95\cdot 10^{-14} \t{ m}/\sqrt{\t{Hz}})^2$), the equivalent bath occupancy of this noise has an exceptionally low value of $n_\t{imp}^\t{ex}\approx 1.0\cdot 10^{-5}$, nearly 44 dB below the value at the SQL.  Encouragingly, the measurement rate associated with this imprecision, $\Gamma_{\t{m}}/16n_\t{imp}^\t{ex}\approx 2\pi\cdot 36$ kHz, is equal to the thermal decoherence rate at an experimentally accessible temperature of $1.3$ K. The more lenient requirements for feedback cooling to $n_\t{m}<1$ (i.e., $\Gamma_\t{meas}<\Gamma_\t{th}/8$) should thus be accessible at 10 K.

The performance of our sensor is limited in practice by constraints on the usable optical power, including photon collection efficiency, photothermal and radiation pressure instabilities, and extraneous sources of measurement back-action, such as heating due to optical absorption. We investigate these constraints by recording the measurement imprecision, $n_{\t{\t{imp}}}$, and the total effective bath occupation, $n_\t{tot}\equiv n_{\t{th}}+n_{\t{ba}}$, as a function of intracavity photon number, $n_\t{c}$ (see S.I.), comparing their product to the uncertainty-limited value, $4\sqrt{n_{\t{\t{imp}}}n_\t{tot}}>1$ (Fig. \ref{fig:imprecision_backaction_figure}). Two considerations are crucial to this investigation.  First, in order to efficiently collect photons from the cavity, it is necessary to increase the taper-cavity coupling rate to $\kappa_\t{ex}\gtrsim\kappa_{0}$, thereby increasing the total cavity decay rate to $\kappa=\kappa_0+\kappa_\t{ex}$.  We operate at a near-critically coupled $(\kappa_\t{ex}= \kappa_0)$ value of $\kappa\approx 2\pi\cdot 0.91$ GHz, thus reducing the single photon cooperativity to $C_0\approx0.31$ in exchange for a higher output coupling efficiency of $\eta_c = (\kappa-\kappa_0)/\kappa\approx0.52$.  Second, in order to minimize $S_x^{\t{\t{imp}}}$, it is necessary to maximize intracavity photon number while mitigating associated dynamic instabilities.  We accomplish this by actively damping the oscillator using radiation pressure feedback.  Feedback is performed by modulating the drive intensity, and therefore the intracavity photon number, of the secondary feedback mode using an electronically amplified and delayed (by $\tau\approx 3\pi/2\Omega_{\t{m}}$) copy of the homodyne photocurrent as an error signal.  
The resulting viscous radiation pressure gives rise to a well-known cooling effect (`cold-damping') \cite{cohadon_cooling_1999,kleckner_sub-kelvin_2006,poggio_feedback_2007,abbott2009observation,li_millikelvin_2011}, reducing the phonon occupancy of the mechanical mode to a mean value of $n_\t{m} \approx n_\t{tot}\Gamma_{\t{m}}/(\Gamma_{\t{m}}+\Gamma_{\t{fb}})$, where $\Gamma_{\t{fb}}$ is the optically-induced damping rate. 
It should be noted that added damping leads to an apparent imprecision, $n'_{\t{\t{imp}}} = n_{\t{\t{imp}}}(\Gamma_{\t{m}}+\Gamma_{\t{fb}})/\Gamma_{\t{m}}$, that differs from the intrinsic value $(\Gamma_{\t{fb}}=0)$.
We here restrict our attention to the latter, noting that the associated cooling preserves the apparent imprecision-back-action product: $n_\t{m}n_{\t{\t{imp}}}'=n_\t{tot}n_{\t{\t{imp}}}$.

Representative measurements of the oscillator's  thermal motion are shown in Fig. \ref{fig:SetupFigure}c.   We determine $n_\t{tot}$ and $n_{\t{\t{imp}}}$ by fitting each noise peak to a Lorentzian with a linewidth of $\Gamma_{\t{eff}} = \Gamma_{\t{m}}+\Gamma_{\t{fb}}+\Gamma_{\t{ba}}$ (including a minor contribution from dynamic back-action, $\Gamma_{\t{ba}}$; see S.I.), a peak amplitude of $S_\omega(\Omega_m) \approx 2n_\t{tot}(\Gamma_{\t{m}}/\Gamma_{\t{eff}})^2S_\omega^{\t{zp}}$, and an offset of $S_\omega^{\t{imp}}=2n_{\t{\t{imp}}}S_\omega^{\t{zp}}$. 
For low intracavity photon number, $n_c\ll n_\t{th}/C_0$, we observe that the effective bath occupation is dominated by thermalization to the cryostat, $n_\t{tot}\approx n_\t{th}$, and that imprecision scales as $n_\t{imp}=(16\xi C_0 n_c)^{-1}$, where $\xi\approx0.23$.   $\xi$ represents the ideality of the measurement, and includes both optical losses and reduction in the cavity transfer function due to mode splitting (see S.I.). Operating with higher input power --- ultimately limited by the onset of parametric instability in higher-order beam modes --- the lowest imprecision we have observed is $n_{\t{imp}}\approx 2.7(\pm0.2)\cdot 10^{-5}$, corresponding to an imprecision $39.7\pm0.3$ dB below that at the SQL.  The associated measurement rate, $\Gamma_{\t{meas}}\approx 2\pi\cdot (13\pm 1)$ kHz, is a factor of $9.2$ lower than the rate of decoherence to the ambient 4.4 K bath, $\Gamma_\t{th}\approx 2\pi\cdot 120$ kHz.  Significantly, this value is within $15\%$ of the requirement for feedback cooling to $n_\t{m}<1$.

For large measurement strengths, quantum measurement back-action (radiation pressure shot noise \cite{purdy_observation_2013}) should in principle exceed the ambient thermal force, scaling as $n_\t{ba}=C_0n_\t{c}$ (see S.I.). As shown in Figure \ref{fig:imprecision_backaction_figure}, our system deviates from this ideal behavior due to extraneous back-action, manifesting as an apparent excess cooperativity, $C_0^\t{ex}\approx 0.56$, and limiting the fractional contribution of quantum back-action to $C_0/(C_0+C_0^\t{ex})\approx 35\%$. Similar behavior for high-order mechanical modes suggests that photo-absorption heating is the cause of this excess back-action, as does our observation that $C_0^\t{ex}$ is markedly higher at lower cryostat temperatures --- consistent with the universal reduction of thermal conductivity in amorphous glasses below 10 K \cite{pohl2002low}.  Combining this extraneous back-action with non-ideal measurement transduction/efficiency, we model the apparent imprecision-back-action product (green curve in Fig. \ref{fig:imprecision_backaction_figure}) as
\begin{equation}
4\sqrt{n_{\t{imp}}n_\t{tot}}=\sqrt{\frac{1}{\xi}\left(1+\frac{n_{\t{th}}}{C_0 n_{\t{c}}}+\frac{C_0^\t{ex}}{C_0}\right)\left(1+\frac{n_{\t{c}}}{n_\t{c}^\t{ex}}\right)},
\label{eq_product}\end{equation}
where $n_\t{c}^\t{ex}\equiv(16\xi C_0 n_\t{imp}^\t{ex})^{-1}$ is the photon number at which extraneous and shot-noise imprecision are equal.  
Operating at $n_\t{c}\approx 5\cdot10^4\ll n_\t{c}^\t{ex}$, we observe a minimum imprecision-back-action product of $4\sqrt{n_\t{imp}n_\t{tot}}\approx 5.0$.  Thus a maximum measurement efficiency of $\Gamma_\t{meas}/\Gamma_\t{tot}\approx 0.040$ is achieved.

To illustrate the utility of this measurement efficiency, we consider what temperature can be reached by increasing the strength of the feedback used to damp the oscillator in Fig. \ref{fig:SetupFigure}c.  The limits of `cold-damping' have been well-studied \cite{courty_quantum_2001,genes_ground-state_2008}. Ignoring back-action due to the weakly driven ($n_c < 100$) feedback optical mode, the effective phonon occupancy of the cooled mechanical mode depends on the balance between coupling to thermal, measurement, and feedback reservoirs at rates $\Gamma_{\t{th}}$, $\Gamma_{\t{m}}n_\t{ba}$, and $g_\t{fb}\Gamma_{\t{m}}n_\t{imp}$, respectively, where $g_\t{fb}\equiv\Gamma_\t{fb}/\Gamma_\t{m}$ is the open loop feedback gain (see S.I.):
\begin{equation}
n_\t{m}+\frac{1}{2}=\frac{1}{1+g_{\t{fb}}}n_\t{tot}+\frac{g_{\t{fb}}^{2}}{1+g_{\t{fb}}}n_{\t{\t{imp}}}\ge  2\sqrt{n_{\t{\t{imp}}}n_\t{tot}}.
\label{eq_cooling}\end{equation}
Note that here $\Gamma_\t{ba}\ll\Gamma_\t{fb}$ has been assumed.  The minimum occupation is achieved for an optimal gain of $g_\t{fb}=\sqrt{n_\t{tot}/n_{\t{\t{imp}}}}$, and corresponds to suppressing the apparent position noise to the imprecision noise floor (cf. yellow curve in Fig. \ref{fig:feedbackcoolingfigure}, inset).  Notably, in the absence of extraneous back-action, $n_\t{m}<1$ requires $n_{\t{imp}}<1/2n_{\t{th}}$.  Results shown in Fig. \ref{fig:imprecision_backaction_figure} suggest that $n_\t{m}\approx2$ should be achievable with our system.

Fig. \ref{fig:feedbackcoolingfigure} shows the result of feedback cooling using a measurement with an imprecision far below that at the SQL. We emphasize that for this demonstration, imprecision was deliberately limited to $n_{\t{imp}}\approx 2.9\cdot 10^{-4}$ in order to reduce uncertainties due to extraneous heating and due to the off-resonant tail of the thermal noise peak at 4.6 MHz (which limits applicability of Eq. \ref{eq_cooling} to effective damping rates of $\Gamma_{\t{eff}}=(1+g_\t{fb})\Gamma_\t{m}\lesssim 2\pi\cdot 200$ kHz).  The feedback gain was controlled by changing the magnitude of the electronic gain, leaving all other parameters (e.g. laser power) unaffected.  Fitting the closed loop noise spectrum (Fig. \ref{fig:feedbackcoolingfigure}, inset) to a standard Lorentzian noise squashing model \cite{poggio_feedback_2007} (see S.I.), we estimate the phonon occupancy of the mechanical mode from the formula $n_\t{m}+0.5 \approx \Gamma_{\t{eff}}\cdot(S_\omega(\Omega_\t{m})+S_\omega^\t{imp})/2S_\omega^{\t{zp}}$, where $S_\omega^\t{imp}$ denotes the off-resonant background. 
Accounting for extraneous back-action, we infer a minimum occupation of $n_\t{m}\approx 5.3\pm0.6$ at an optimal damping rate of $\Gamma_{\t{eff}}\approx 2\pi\cdot 52$ kHz, corresponding to a ground state probability of $1/(1+n_\t{m})\approx 16\%$.  The value agrees well with the prediction based on Eq. \ref{eq_cooling} and Fig. \ref{fig:imprecision_backaction_figure}.  Notably, for larger feedback strengths, shot noise squashing \cite{wiseman2009quantum,cohadon_cooling_1999} leads to an apparent reduction of $n_\t{imp}$, while $n_\t{m}$ physically increases. 

Collectively, our results establish new benchmarks for linear measurement and control of a micromechanical oscillator.  The enabling advance is a position sensor capable of monitoring the oscillator's displacement with an imprecision $39.7\pm0.3$ dB below that at the SQL, a 100-fold improvement over results reported to date \cite{teufel_nanomechanical_2009,anetsberger_measuring_2010,westphal_interferometer_2012}, combined with an imprecision-back-action product within a factor of $5.0$ of the uncertainty limit. 
Achieving this sensitivity requires the use of a small mass, high-Q mechanical oscillator operating in a cryogenic environment.  For our system, a 4.3 MHz nanomechanical beam oscillator operating at 4.4 K, the achieved imprecision corresponds to the ability to resolve the oscillator's zero-point motion within an order of magnitude of its intrinsic thermal decoherence rate, $\Gamma_\t{meas}/\Gamma_\t{th}\approx 0.11$, and with a total measurement efficiency of $\Gamma_\t{meas}/\Gamma_\t{tot}\approx$ 0.040.  Taking advantage of this efficiency, we show that traditional radiation pressure cold-damping \cite{cohadon_cooling_1999} can be used to cool the oscillator to a mean phonon occupancy of $5.3\pm0.6$; this represents a 40-fold improvement over previous active feedback cooling applied to solid-state mechanical oscillators \cite{poggio_feedback_2007,corbitt_optical_2007,kleckner_sub-kelvin_2006,li_millikelvin_2011,abbott2009observation}, and invites comparison \cite{genes_ground-state_2008,hamerly_advantages_2012,jacobs_when_2014} to the success of coherent feedback (i.e. sideband) cooling in cavity optomechanics \cite{chan_laser_2011,teufel_sideband_2011}. 
With moderate reduction of extraneous back-action, we anticipate that $n_\t{m}<1$ should be possible. Looking forward, high efficiency optomechanical sensors open the door to a variety of measurement-based feedback applications, notably back-action evasion \cite{wiseman_using_1995,heidmann2004beating} and mechanical squeezing \cite{szorkovszky2011mechanical}.  
{
\newcommand{\nocontentsline}[3]{}
\renewcommand{\addcontentsline}[2][]{\nocontentsline#1{#2}}
\begin{acknowledgements}
This research was carried out with support from an ERC Advanced Grant and from the Swiss National Science Foundation, through grants from QCIT and NCCR of Quantum Engineering. N.P. and D.J.W. gratefully acknowledge support from the European Commission through Marie Skłodowska-Curie Fellowships: IEF project 303029 and IIF project 331985, respectively.
\end{acknowledgements}


%

}

\clearpage


\onecolumngrid
\normalsize

\begin{center}
\Large{
\textbf{Supplementary information for\\ 
	``Measurement and control of a mechanical oscillator at its thermal decoherence rate''}}
\end{center}
\vspace{-0.2in}

\setcounter{tocdepth}{3}
\tableofcontents


\setcounter{equation}{0}
\setcounter{figure}{0}
\setcounter{table}{0}


\vfill \pagebreak
\section*{Glossary of important variables}

\begin{table}[h!]
\begin{tabular}{|c|p{12cm}|}
\hline
\multicolumn{2}{|c|}{\textbf{Introduced in \secref{sec:theoryFeedback}}} \\
\hline $x,p$							& position and momentum operator of the mechanical oscillator \\
\hline $x_\t{zp},p_\t{zp}$			& $\sqrt{\mathrm{variance}}$ of the position and momentum in the ground state \\
\hline $y$							& position estimate (apparent position inferred from measurement) \\
\hline $x_\t{imp}$					& imprecision of position estimate \\
\hline $\Omega_\t{m},\Gamma_\t{m}$	& resonant frequency and damping rate of the oscillator \\
\hline $m$							& effective mass of the (extended) elastic oscillator mode \\
\hline $\chi_\t{m}$					& intrinsic susceptibility of the oscillator position to an external force \\
\hline $T$							& temperature of the ambient thermal environment \\
\hline $F_\t{th}$					& Langevin force associated with the ambient thermal environment \\
\hline $F_\t{ba},F_\t{ba,th}$		& Measurement back-action force, Langevin force associated with measurement noise \\
\hline $F_\t{fb},F_\t{fb,th}$		& Feedback force, Langevin force associated with feedback noise \\
\hline $\chi_\t{fb},\chi_\t{ba}$	& susceptibility of the linear feedback network, and back-action \\

\hline $\chi_\t{eff},\Gamma_\t{eff}$ & effective susceptibility and damping rate of the oscillator in the presence of feedback and/or back-action \\
\hline $n_\t{th},n_\t{ba},n_\t{fb}$	 & effective thermal occupancy of the ambient thermal bath, measurement (`back-action') noise reservoir, and feedback noise reservoir.
\\
\hline $n_\t{imp}$					& position imprecision referred to an effective thermal occupation \\ 
\hline $n_\t{m},n_\t{m,min}$	& mean phonon occupancy of the oscillator in the presence of feedback, minimal possible occupation \\
\hline $g_\t{fb},g_\t{fb,opt}$		& feedback gain, optimal value of feedback gain \\
\hline $\Gamma_\t{th}$				& thermal decoherence rate ($=n_\t{th}\Gamma_\t{m}$) \\
\hline
\multicolumn{2}{|c|}{\textbf{Introduced in \secref{sec:theoryOptomechanics}}} \\
\hline $\omega_\t{c},\omega_\ell,\lambda$	& cavity resonance frequency, carrier frequency of optical input, wavelength of optical carrier
													($\lambda = 2\pi c/\omega_\ell$, where $c$ is the speed of light) \\
\hline $\Delta_0, \Delta$ 			& bare and renormalized laser-cavity detuning $(\Delta_0 =\omega_\ell -\omega_c)$ \\ 
\hline $\kappa_0, \kappa_\t{ex},\kappa$ & intrinsic, coupling-induced, and total cavity decay rates $(\kappa = \kappa_0 + \kappa_{ex})$  \\ 
\hline $\eta_\t{c}$					& cavity coupling efficiency $(= \kappa_\t{ex}/\kappa)$ \\
\hline $\gamma$ 					& cavity mode splitting \\ 
\hline $a_{\pm},c$ 					& intracavity field operator for the measurement ($a$) and feedback ($c$) optical mode, in a frame rotating at $\omega_\ell$; 
										$+(-)$ refers to field propagating in (counter) the direction of injected power  \\ 
\hline $n_\pm$ 						& steady state intracavity photon numbers \\
\hline $s_\t{in}^\pm$				& field operators for modeling for the injected traveling wave, in a frame rotating at $\omega_\ell$ \\ 
\hline $\delta s_\t{in}^\pm,\delta s_\t{vac}^\pm$ 	& white noise operators modeling the vaccuum fluctuations coupled in via the travelling wave inputs (`in'),
										and the intrinsic decay channels of the cavity (`vac') \\ 
\hline $\chi_a^{(\gamma)}$			& susceptibility of the optical field to cavity input noises, in the presence of splitting $\gamma$ \\
\hline $g_0$ 						& vacuum optomechanical coupling rate  \\ 
\hline $u$ 							& normalized mechanical position operator $(=x/x_\t{zp})$  \\ 
\hline $\delta f_\t{th},\xi_\t{th}$	& normalized thermal force and associated white noise operator \\
\hline $f_\t{ba},\delta f_\t{ba}$	& dynamical and stochastic components of the normalized radiation pressure force from the measurement mode \\
\hline $f_\t{fb}$					& normalized radiation pressure feedback force\\
\hline $\Omega_\t{ba},\Gamma_\t{ba}$ & mechanical frequency and damping rate renormalized by dynamic back-action \\
\hline $C_0$						& single-photon cooperativity ($= 4g_0^2/\kappa\Gamma_\t{m}$) \\
\hline
\multicolumn{2}{|c|}{\textbf{Introduced in \secref{sec:Experimental}}} \\
\hline $G$							& optomechanical coupling expressed as a cavity frequency pull parameter 
										$(=\partial \omega_c/\partial x = g_0/x_\t{zp})$ \\
\hline
\end{tabular} 
\end{table}

\section*{Note on convention}

For any variable $X$, $\bar{X}$ denotes its classical steady state value, and $\delta X(t) = X(t) -\bar{X}$,
the fluctuation from that steady state. We define the Fourier transform of an operator $X(t)$ by
\begin{equation}
	\tilde{X}(\Omega) \eqdef \int_{-\infty}^{+\infty} X(t)e^{i\Omega t}\, dt.
\end{equation}
Following standard definition \cite{Clerk10}, we employ the symmetrized spectral density,
\begin{equation}
	\bar{S}_{XX}(\Omega) \eqdef \int_{-\infty}^{\infty} \frac{1}{2}
		\avg{\delta X(t)\, \delta X(0)+ \delta X(0)\, \delta X(t)}e^{i\Omega t}\, dt,
\end{equation} 
to describe the spectral distribution of the variance of the operator-valued process,
$\delta X(t)$.  The single-sided spectral density is then given by (strictly for $\Omega \geq 0$)
\begin{equation}
	S_{X}(\Omega) = 2\, \bar{S}_{XX}(\Omega).
\end{equation}
To make contact with experiment and with the main text, we hereafter adopt the single-sided convention unless otherwise necessary for clarity.


\section{Theory of feedback cooling of a harmonic oscillator}\label{sec:theoryFeedback}

Consider a harmonic oscillator, whose motion is described by position coordinate $x(t)$, moving in a harmonic potential
of frequency $\Omega_m$. We specialize to a case in which the oscillator is subject to three stochastic forces: a
thermal force ($F_\t{th}$) associated with the ambient environment, a `back-action' force ($F_\t{ba}$) associated with
the oscillator's coupling to a measurement device, and a feedback force ($F_\t{fb}$) that controls the oscillator.  The
dynamics of this system are described by the Langevin equation \footnote{we adopt a Brownian motion description  of the
oscillator position, as opposed to a Lindblad one, in spite of the well-known non-positivity of the former \cite{Lind76,GnutHaak96};
this choice is justified by their mutual concordance at sufficiently large temperatures 
$T \gtrsim \frac{\hbar \Gamma_m}{k_B} = \mathcal{O}(10^{-10}\, \t{K})$ \cite{Haak85,JacTitt09}},
\begin{equation}\label{eq:eomclassical}
\begin{split} 
	& m\left(\ddot{x}+\Gamma_\t{m} \dot{x} + \Omega_\t{m}^2 x\right)  = F_\t{th} + F_\t{ba} + F_\t{fb}\\
	\Rightarrow\;\; & \underbrace{\left( \Omega_\t{m}^2 -\Omega^2 -i\Omega \Gamma_\t{m} \right)}_{\eqdef 
		\chi_\t{m} (\Omega)^{-1}} \tilde{x} = m^{-1}\left( \tilde{F}_\t{th} + \tilde{F}_\t{ba} + \tilde{F}_\t{fb} \right).
\end{split} 
\end{equation}
where $\chi_\t{m}$ is the intrinsic mechanical susceptibility (note that, for notational convenience in 
section \ref{sec:theoryOptomechanics}, we have scaled the conventional expression for $\chi_\t{m}$ by $m$).

We adopt the following model for the back-action and feedback forces:
\begin{subequations}
\begin{align}\label{eq:feedback}
	\tilde{F}_\t{ba} &= -m \chi_\t{ba}(\Omega)^{-1}\,\tilde{x}+\tilde{F}_\t{ba,th}\\
	\tilde{F}_\t{fb} &= -m \chi_\t{fb}(\Omega)^{-1}\,\tilde{y}+\tilde{F}_\t{fb,th}.
\end{align}
\end{subequations}
Each force has two components: a `dynamic' component, characterized by a linear susceptibility, that contains
correlations with the oscillator's position, and an effective thermal component.  Notably, the dynamic component of the
feedback force is linear in an apparent (measured) position, $\tilde{y}=\tilde{x}+\tilde{x}_\t{imp}$, where 
$\tilde{x}_\t{imp}$ is the measurement imprecision.  Hereafter, for simplicity, we neglect the dynamic portion of the
back-action force.  We revisit this approximation in detail in section \ref{sec:probeDBA}.

\subsection{Optimal feedback cooling}

In feedback cooling, we are interested in minimizing the mean phonon occupancy of the oscillator, viz.
\begin{equation}
	\min_{\chi_\t{fb}}\, (2n_\t{m}+1) = 
	\min_{\chi_\t{fb}}\, \sqrt{\avg{\frac{x^2}{x_\t{zp}^2}} \avg{\frac{p^2}{p_\t{zp}^2}}
		-\avg{\frac{xp+px}{2x_\t{zp}p_\t{zp}}}},
\end{equation}
where $x_\t{zp}^2 = \frac{\hbar}{2m\Omega_\t{m}}$, and $x_\t{zp}p_\t{zp}=\frac{\hbar}{2}$.
For a thermal state, the constraint reduces to
\begin{equation}\label{eq:cost}
	\min_{\chi_\t{fb}}\, \frac{\avg{x^2}}{x_\t{zp}^2}.
\end{equation}

The problem described herein --- that of a linear system driven by weak-stationary Gaussian noise 
(\eqref{eq:eomclassical}), and controlled by linear measurement and actuation (\eqref{eq:feedback}), with the aim of 
minimizing a quadratic cost function (\eqref{eq:cost}) --- is an archetype of the \emph{linear quadratic 
gaussian} (LQG) paradigm of classical control theory \cite{Wien49,AthLQG71,Kail80}. 
Recently, such problems have been formalized and studied in the quantum mechanical context \cite{James08,NurdinLQG09}.

The optimal feedback filter, $\chi_\t{fb}$, can be solved for exactly in our case. From \eqref{eq:eomclassical} and
\eqref{eq:feedback} (and neglecting $\chi_\t{ba}$), the oscillator position and measurement record are
\begin{equation}\label{eq:eomxmeas}
\begin{split}
		& \left( \chi_\t{m}^{-1} +\chi_\t{fb}^{-1}\right) \tilde{x} = m^{-1}\left(
			\tilde{F}_\t{th}+\tilde{F}_\t{ba,th}+\tilde{F}_\t{fb,th}\right) - \chi_\t{fb}^{-1} \tilde{x}_\t{imp} \\
		& \left( \chi_\t{m}^{-1} +\chi_\t{fb}^{-1}\right) \tilde{y} = m^{-1}\left( \tilde{F}_\t{th}+\tilde{F}_\t{ba,th}+\tilde{F}_\t{fb,th}\right)
			 +\chi_\t{m}^{-1} \tilde{x}_\t{imp}.
\end{split}
\end{equation}
Identifying the effective mechanical susceptibility, $\chi_\t{eff}^{-1} \eqdef \chi_\t{m}^{-1} + \chi_\t{fb}^{-1}$ and total 
effective thermal force, $S_{F}^\t{tot} =S_{F}^\t{th}+S_{F}^\t{ba}+S_{F}^\t{fb,th}$, \eqref{eq:eomxmeas} implies
\begin{equation}\label{eq:SxxSyy}
\begin{split}
	& S_{x}(\Omega) = \vert \chi_\t{eff}\vert^2 \left(m^{-2}
		S_{F}^\t{tot}+ \vert \chi_\t{fb}\vert^{-2} S_{x}^\t{imp} \right) \\
	& S_{y}(\Omega) = \vert \chi_\t{eff}\vert^2 \left(m^{-2}S_{F}^\t{tot} + \vert \chi_\t{m}\vert^{-2} S_{x}^\t{imp} \right).
\end{split}
\end{equation}
for the (single-sided) spectral density of the position and measurement records.

The LQG problem can be stated concretely as (using double-sided spectra temporarily)
\begin{equation}\label{eq:feedbacktarget}
	\min_{\chi_\t{fb}}\, \int_{-\infty}^{\infty} \bar{S}_{xx}(\Omega)\, \frac{d\Omega}{2\pi}.
\end{equation}
The solution to this variational problem is given by the Euler-Lagrange equation,
\begin{equation}
	\frac{\mathscr{D}\bar{S}_{xx}}{\mathscr{D}\chi_\t{fb}} = 0,
\end{equation}
where $\mathscr{D}$ stands for the variational (Gateaux) derivative.
This is most effectively solved in terms of the magnitude $\vert \chi_\t{fb}\vert$ and phase $\phi_\t{fb}=\arg \chi_\t{fb}$ of the optimal filter; 
resulting in the solution,
\begin{equation}\label{eq:optimal}
\begin{split}
	& \phi_\t{fb}(\Omega) = \arctan \frac{\im \chi_\t{m}}{\re \chi_\t{m}} \\
	& \vert \chi_\t{fb}(\Omega)\vert\,\bar{S}_{FF}^\t{tot} = \vert \chi_\t{m}(\Omega)\vert^{-1}\,\bar{S}_{xx}^\t{imp}.
\end{split}
\end{equation}

\subsection{Practical feedback cooling}\label{sec:practicalfeedback}

The optimal feedback phase near resonance is
\begin{equation}
	\phi_\t{fb}(\Omega) = \arctan \frac{\Omega \Gamma_m}{\Omega_\t{m}^2 -\Omega^2}
		\approx \pm\frac{\pi}{2}+2\frac{\Omega-\Omega_\t{m}}{\Gamma_\t{m}}.
\end{equation}
In practice, it is easiest to implement $\phi_\t{fb}=\pi/2$ across the mechanical oscillator bandwidth, and
choose $\vert \chi_\t{fb}\vert^{-1} \propto \Omega$, i.e.,
\begin{equation}
	\chi_\t{fb}(\Omega)^{-1} = -i\Omega \Gamma_\t{fb}(\Omega),
\end{equation}
where ideally the feedback gain $\Gamma_\t{fb}(\Omega) = g_\t{fb}\Gamma_m$, with $g_\t{fb}$ the dimensionless 
gain of the filter. The ensuing effective susceptibility,
\begin{equation}
	\chi_\t{eff}^{-1} \eqdef \chi_\t{m}^{-1} + \chi_\t{fb}^{-1} = \Omega_\t{m}^2 -\Omega^2 
		+i\Omega \underbrace{\Gamma_\t{m}(1+g_\t{fb})}_{\eqdef \Gamma_\t{eff}}, 
\end{equation}
is characterized by a modified damping rate, $\Gamma_\t{eff}$.  

To see how this damping leads to cooling, we reconsider the three components of the thermal environment: (1) an ambient reservoir with 
which the oscillator equilibrates, (2) a reservoir constituted by stochastic measurement back-action, and (3) a reservoir constituted 
by stochastic fluctuations of the feedback actuator. For a high-Q oscillator, each reservoir can be assigned a thermal noise equivalent 
occupation: $n_\t{th}$, $n_\t{ba}$ and $n_\t{fb}$ respectively, where $n_\t{th}=\tfrac{1}{2}\coth\left(\hbar \Omega_\t{m}/2k_B T\right)$ 
in terms of the ambient bath temperature, $T$. Thus the total effective thermal force may be expressed:
\begin{equation}\label{eq:SFFdef}
 S_\t{F}^\t{tot}(\Omega)= \left(n_\t{th}+n_\t{ba}+n_\t{fb}+\tfrac{1}{2}\right)\cdot m^2\vert\chi_\t{m}(\Omega_m)\vert^{-2}\cdot 2S_{x}^\t{zp}(\Omega_\t{m}).
\end{equation}
where we have introduced for convenience the (peak) position spectral density in the ground state:
\begin{equation}\label{eq:Szp}
	S_{x}^\t{zp}(\Omega_\t{m}) = \frac{4 x_\t{zp}^2}{\Gamma_\t{m}}.
\end{equation} 

We further introduce the imprecision quanta, $n_\t{imp}$, as the apparent thermal occupation associated with noise in the measurement:
\begin{equation}\label{eq:Simpdef}
\begin{split} 
	S_x^\t{imp}(\Omega) &= n_\t{imp} \cdot 
		2S_x^\t{zp}(\Omega_\t{m}).
\end{split} 
\end{equation}

Thus the spectra of physical position and the measurement record, \eqref{eq:SxxSyy}, are given by
\begin{equation}\label{eq:SxSxmeas}
\begin{split}
	\frac{S_{x}(\Omega)}{2S_{x}^\t{zp}(\Omega_\t{m})} &= 
		\frac{(n_\t{th}+n_\t{ba}+n_\t{fb}+\tfrac{1}{2})\Omega_\t{m}^2 \Gamma_\t{m}^2 
		+ n_\t{imp}\, g_\t{fb}^2 \Omega^2 \Gamma_\t{m}^2}
		{(\Omega_\t{m}^2 -\Omega^2)^2 +\Omega^2 \Gamma_\t{eff}^2} \\
	\frac{S_{y}(\Omega)}{2S_{x}^\t{zp}(\Omega_\t{m})} &= 
		\frac{(n_\t{th}+n_\t{ba}+n_\t{fb}+\tfrac{1}{2})\Omega_\t{m}^2 \Gamma_\t{m}^2 + 
		n_\t{imp}\, \left((\Omega_\t{m}^2 -\Omega^2)^2 + \Omega^2 \Gamma_\t{m}^2\right)}
		{(\Omega_\t{m}^2 -\Omega^2)^2 +\Omega^2 \Gamma_\t{eff}^2}.
\end{split}
\end{equation}
The mean phonon occupancy of the cooled oscillator is then given by
\begin{equation}\label{eq:nEff}
\begin{split} 
	& 2n_\t{m}+1 = \int_{0}^{\infty}\frac{S_x(\Omega)}{x_\t{zp}^2}\, \frac{d\Omega}{2\pi} \\
	\Rightarrow\;\; & n_\t{m} = \frac{(n_\t{th}+n_\t{ba}+n_\t{fb}+\tfrac{1}{2})+n_\t{imp}g_\t{fb}^2}{1+g_\t{fb}}-\frac{1}{2}.
\end{split} 
\end{equation}
In the relevant limit of $n_\t{th}\gg\tfrac{1}{2}$, a minimum of
\begin{equation}\label{eq:nEffMin}
\begin{split}
	n_\t{m,min} &\approx 2\sqrt{(n_\t{th}+n_\t{ba}+n_\t{fb})n_\t{imp}}-\frac{1}{2}\\&
	\approx\frac{1}{2\hbar}\sqrt{S_F^\t{tot}(\Omega_\t{m})S_x^\t{imp}(\Omega_\t{m})}-\frac{1}{2}
\end{split}
\end{equation}
is attained at an optimal gain of
\begin{equation}
	g_\t{fb,opt} 
	\approx \sqrt{\frac{n_\t{th}+n_\t{ba}+n_\t{fb}}{n_\t{imp}}},
\end{equation}
as anticipated by \eqref{eq:optimal}.

In particular, for the experimentally relevant case of $n_\t{th}\gg n_\t{fb}$, the conventional condition for ground state cooling, $n_\t{m} < 1$, translates to
\begin{equation}\label{eq:nimpGS1}
	n_\t{imp} < \frac{9}{16}(n_\t{th}+n_\t{ba})^{-1}.
\end{equation}

Finally, in the regime where feedback cooling is strong ($g_\t{fb} \gg 1$) and quantum-limited ($n_\t{fb}=0$), intuition can be 
garnered by noticing that \eqref{eq:nEff} can be expressed as the detailed balance condition, 
$\left(n_\t{m}+\tfrac{1}{2}\right) \Gamma_\t{eff} = (n_\t{th}+n_\t{ba})\Gamma_\t{m} + n_\t{imp}\Gamma_\t{fb}$. 
This suggests that cooling as affected by feedback can be understood as a thermodynamic process which proceeds by the
reduction of entropy of the mechanical oscillator to a level ultimately set by the entropy due to the imperfect
estimation of the mechanical position. 

\subsection{Limits due to stochastic back-action}

In section \ref{sec:SQL}, it is shown that stochastic back-action associated with a cavity-optomechanical position measurement is bound by 
the imprecision-back-action product: $\hbar^2S_x^\t{imp}S_F^\t{ba}=16n_\t{imp}n_\t{ba}\ge 1$.  Imposing this limit, \eqref{eq:nimpGS1} implies that a 
necessary condition for ground-state cooling is
\begin{equation}\label{eq:nimpGS2}
	n_\t{imp} < (2n_\t{th})^{-1}.
\end{equation}
Notably, from \eqref{eq:Simpdef}, the associated condition on the measurement imprecision becomes
\begin{equation}\label{eq:impGScooling}
S_{x}^\t{imp}< \frac{S_{x}^\t{zp}}{n_\t{th}}=\frac{4 x_\t{zp}^2}{n_\t{th}\Gamma_\t{m}}=\frac{4 x_\t{zp}^2}{\Gamma_\t{th}},
\end{equation}
where $\Gamma_\t{th}\eqdef \Gamma_\t{m} n_\t{th}$ is the thermal decoherence rate.  Notably \eqref{eq:impGScooling} 
corresponds to an imprecision $n_\t{th}/2$  times below that at the standard quantum limit (\eqref{eq:SQL}), or equivalently, 
as a measurement rate \cite{Makh01,Clerk10}
\begin{equation}
\Gamma_\t{meas}\eqdef\frac{x_\t{zp}^2}{2 S_x^\t{zp}}=\frac{\Gamma_\t{m}}{16n_\t{imp}}>\frac{\Gamma_\t{th}}{8}
\end{equation}


\vfill \pagebreak 
\section{Readout and feedback using a cavity}\label{sec:theoryOptomechanics}

In our system, the mechanical oscillator is dispersively coupled to an optical cavity mode. The cavity field exerts a 
radiation pressure force on the oscillator; the unitary nature of this interaction affects a phase shift of the cavity 
field commensurate with the amplitude of mechanical motion. \\

\begin{figure}[h!]
	\includegraphics[scale=0.5]{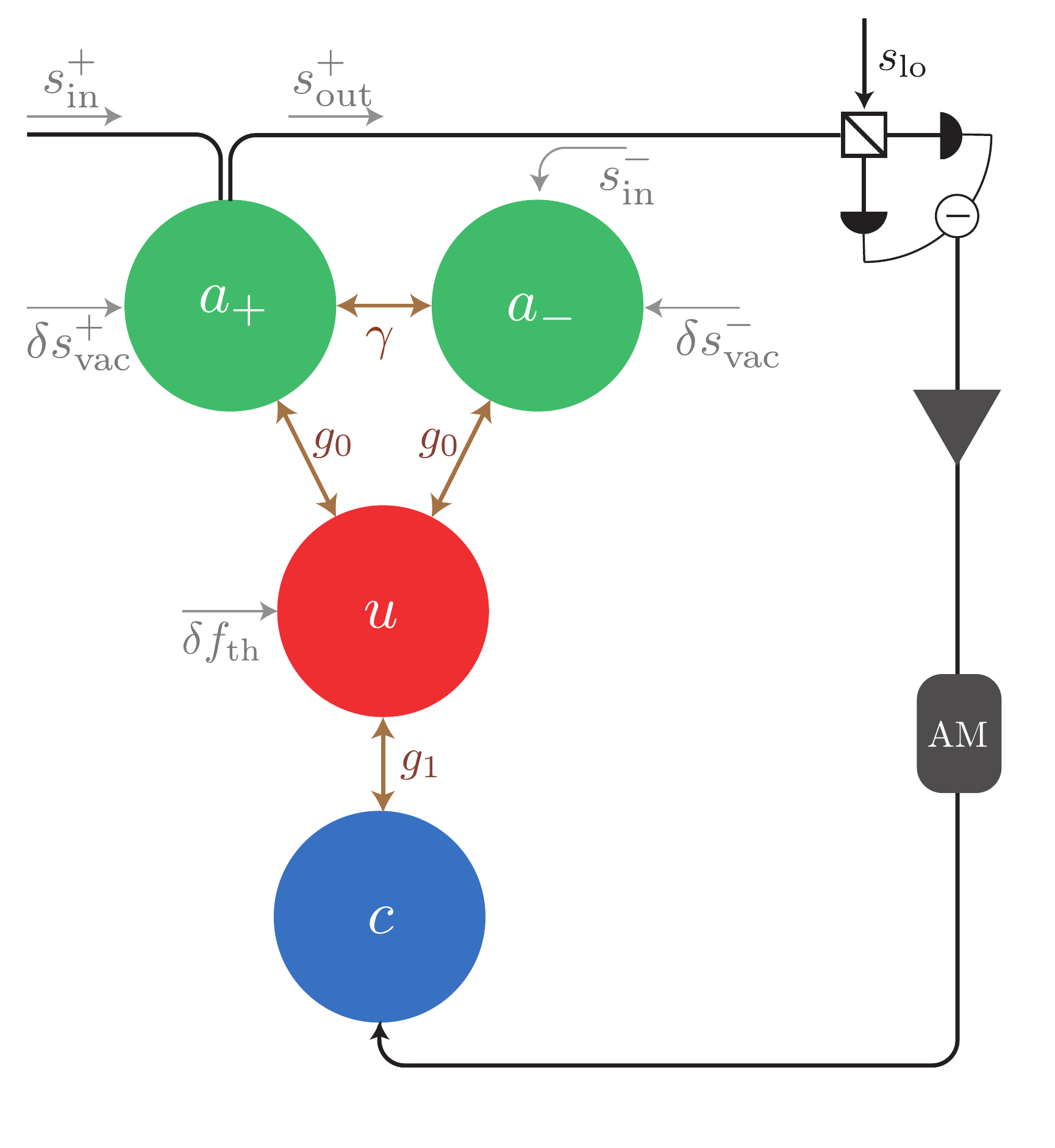}
	\caption{\label{fig:theorySchematic} Schematic of the relevant input, output and couplings between the various
		subsystems in the experiment.}
\end{figure}

We adopt the following set of coupled Langevin equations to model the dynamics of the cavity mode (characterized by the 
slowly varying amplitude of the intracavity field, $a$) and the mechanical mode (characterized by its normalized position, $u\eqdef x/x_\t{zp}$):
\begin{subequations} \label{eq:OMeqofmotion}
\begin{equation}\label{eq:OMeqofmotion_optics}
\begin{split}
	 &\dot{a}_+ = \left(i\Delta_0 -\frac{\kappa}{2}\right) a_+ + \frac{i\gamma}{2}a_- + i g_0 u a_+
			+ \sqrt{\eta_\t{c} \kappa}\, s_\t{in}^{+} + \sqrt{(1-\eta_\t{c})\kappa}\, \delta s_\t{vac}^{+} \\
	 &\dot{a}_- = \left(i\Delta_0 -\frac{\kappa}{2}\right) a_- + \frac{i\gamma}{2}a_+ + i g_0 u a_-
			+ \sqrt{\eta_\t{c} \kappa}\, s_\t{in}^{-} + \sqrt{(1-\eta_\t{c})\kappa}\, \delta s_\t{vac}^{-}
\end{split}\end{equation}
\begin{equation}\label{eq:OMeqofmotion_mechanics}
	 \ddot{u}+\Gamma_\t{m} \dot{u} + \Omega_\t{m}^2 u = \delta f_\t{th} + f_\t{ba} + f_\t{fb}.
\end{equation}
\end{subequations} 

Notably, in \eqref{eq:OMeqofmotion_optics} we use a two-mode model to describe the microdisk cavity.  Subscripts $+$ and
$-$ refer to whispering gallery modes propagating along (`clockwise') and against (`counter-clockwise') the conventional
direction ($+$) of the injected field, respectively.  The two modes are coupled at a rate $\gamma$ by scattering centers
\cite{kippenberg2002}, leading to a characteristic splitting of the optical resonance (cf. \eqref{eq:steadyState}
and \secref{sec:cavity mode splitting}). Motivated by the geometrical nature of the interaction, we assume that both
modes share a common vacuum optomechanical coupling rate, $g_0$.  

Light is physically coupled to the microdisk cavity using an optical fiber \cite{kippenberg2002}. In Fig.
\ref{fig:theorySchematic} and \eqref{eq:OMeqofmotion_optics}, we model this coupler as a two port waveguide. Fields
entering(exiting) the `clockwise' port, $s^{+}_\t{in(out)}$, couple directly to the clockwise cavity mode. Fields
entering(exiting) the `counter-clockwise' port, $s^{-}_\t{in(out)}$, couple directly to the counter-clockwise mode.  The
cavity-waveguide coupling rate is $\kappa_\t{ex}=\eta_c\kappa$, where $\kappa = \kappa_\t{ex}+\kappa_\t{0}$ is the total
cavity decay rate and $\kappa_\t{0}$ is the intrinsic cavity decay rate. In addition, each cavity mode is driven through
its intrinsic decay channel by a vacuum state with amplitude $\delta s_\t{vac}^{\pm}$. Input field amplitudes are here
normalized so that $P_\t{in}^{\pm} = \hbar\omega_{\ell}^{\pm} |s^{\pm}_\t{in}|^2$ is the injected power.
$\Delta_0=\omega_\t{c}-\omega_{\ell}^{\pm}$ denotes the detuning of the drive field carrier frequency,
$\omega_{\ell}^{\pm}$, from the center frequency of the optical mode doublet, $\omega_{c}$.

The dynamics of the mechanical oscillator are governed by \eqref{eq:OMeqofmotion_mechanics}. Note that owing to the
dimensionless form of $u$, generalized forces $f_\t{th,ba,fb}$ have dimensions of $(\mathrm{time})^{-2}$; the actual
forces (cf. \secref{sec:theoryFeedback}), in units of Newtons, are given by $F_i=m x_\t{zp}\, f_i$ ($i \in
\{\t{th,opt,fb}\}$). Using this convention, the thermal Langevin force is given by
\begin{equation}\label{eq:generalizedlangevin}
	\delta f_\t{th}=\Omega_\t{m} \Gamma_\t{m} \sqrt{2(2n_{th}+1)}\, \xi_\t{th}, \;\;\;\; \t{where} \;\;\;
	n_\t{th}=\frac{1}{2}\coth\left(\frac{\hbar \Omega_\t{m}}{2k_B T}\right)
\end{equation}
and $\xi_\t{th}$ the is unit variance white noise process modeling the bath fluctuations. 


We model the measurement back-action force as the radiation pressure imparted by the excited mode doublet 
$a_{\pm}$: 
\begin{equation}
	f_\t{ba} = \Omega_\t{m} g_0 (a_+^\dagger a_{+} + a_{-}^\dagger a_{-}).
\end{equation}

We likewise model the feedback force as the radiation pressure imparted by an independent, auxiliary cavity mode with
amplitude $c$ and optomechanical couplng rate $g_1$: 
\begin{equation}
	f_\t{fb} = \Omega_\t{m} g_1\, c^\dagger c.
\end{equation}

In the following treatment, both optical modes are driven by optical fields entering the clockwise port of the optical
fiber.  The field driving mode doublet $a_{\pm}$ is refered to as the `sensor' field.  The field driving mode $c$ is
referred to as the `feedback' field.  The counter-clockwise port of the optical fiber is used to monitor the transmitted
sensor field, but is otherwise left open.  We hereafter confine our attention to the back-action force associated with
the cavity mode $a_{\pm}$.\\

\noindent \textbf{Steady state.} When the cavity is excited by the sensor field, the static component of the ensuing
radiation pressure force displaces the oscillator to a new steady-state position, $\bar{u}$, and leads to a
renormalization of the laser-cavity detuning to $\Delta = \Delta_0 +g_0 \bar{u}$.  In practice the frequency of the
sensor field is stabilized so that $\Delta = 0$.  In this case the steady state intracavity field amplitude ($\bar{a}$)
and oscillator position are given by
\begin{equation}\label{eq:steadyState}
\begin{split}
	& \bar{a}_+ = \sqrt{n_+}, \;\; \bar{a}_- = i \sqrt{n_-} \;\; \mathrm{and}\;\; 
	\bar{u} = \frac{g_0}{\Omega_\t{m}}(n_+ + n_-), \\
	\mathrm{where}\;\; 
	& n_+ = \frac{4\eta_\t{c}}{\kappa}\frac{P_\t{in}^{+}/\hbar \omega_\t{c}}{(1+\gamma^2/\kappa^2)^2} \;\;
	\mathrm{and}\;\; n_- = \left(\frac{\gamma}{\kappa}\right)^2 n_+.
\end{split} 
\end{equation}
denote the mean intracavity photon number of the clockwise and counter-clockwise modes, respectively. Note that when
describing Fig. 2 in the main text, we associate the intravity photon number with that of the clockwise mode, i.e.
$n_\t{c}=n_+$.

Splitting of the cavity resonance can be observed spectroscopically in the normalized steady state transmission.  Using
the input-output relation $\bar{s}^+_\t{out}=\bar{s}^+_\t{in}-\sqrt{\eta_\t{c} \kappa}\, \bar{a}_+$ gives
\begin{equation}\label{eq:transmission}
	\left\vert \frac{\bar{s}^+_\t{out}}{\bar{s}^+_\t{in}} \right\vert^2 = \frac{P_\t{out}^{+}}{P_\t{in}^{+}} = 1-\eta_\t{c}
\kappa^2 
	\frac{\left(\Delta^2 +(\gamma/2)^2+(\kappa/2)^2\right)-\eta_\t{c}\left(\Delta^2+(\kappa/2)^2\right)}
	{\left(\Delta^2 -(\kappa/2)^2-(\gamma/2)^2\right)^2},
\end{equation} 
which is used in \secref{sec:cavity mode splitting}.\\

\noindent \textbf{Fluctuations.} Fluctuations of the cavity field, $\delta a = a - \bar{a}$, and the mechanical
position, $\delta u = u-\bar{u}$, are coupled according to \eqref{eq:OMeqofmotion}.  To first order:
\begin{subequations}\label{eq:fluctuations_probe}
\begin{equation}
	\delta \dot{a}_\pm = \left(i\Delta-\frac{\kappa}{2}\right)\delta a_\pm + \frac{i\gamma}{2}\delta a_\mp
		+ ig_0 \bar{a}_\pm\, \delta u + + \sqrt{\eta_\t{c} \kappa}\, \delta s_\t{in}^{-}\pm 
		+ \sqrt{(1-\eta_\t{c})\kappa}\, \delta s_\t{vac}^\pm
\end{equation}
\begin{equation}\label{eq:fluctuations_mechanics_time}
	\delta \ddot{u}+\Gamma_\t{m} \delta \dot{u}+\Omega_\t{m}^2 \delta u = \delta f_\t{th}+\delta f_\t{fb}+
		g_0\Omega_\t{m} \sum_{j=\pm} (\bar{a}_j \delta a_j^\dagger + \bar{a}_j^* \delta a_j).
\end{equation}
\end{subequations}
The ensuing radiation pressure force fluctuations
\begin{equation}\label{eq:backaction}
\delta{f}_\t{ba} = 
		g_0\Omega_\t{m} \sum_{j=\pm} (\bar{a}_j \delta a_j^\dagger + \bar{a}_j^* \delta a_j)
\end{equation}
contain both a dynamic and stochastic component, as detailed in section \ref{sec:probeDBA} and \ref{sec:probeQBA},
respectively.
   
Taking the Fourier transforms of \eqref{eq:fluctuations_probe} recasts the optomechanical interaction in terms of
optical(mechanical) susceptibilities, $\chi_\t{a(m)}$:
\begin{subequations}\label{eq:fluctuations_probe_fourier}
	\begin{equation}
	\begin{aligned}
		\chi_\t{a}^{(\gamma)}(\Omega)^{-1}\, \delta \tilde{a}_\pm =\,
				& ig_0 \left(\bar{a}_\pm +\frac{i\gamma}{2}\chi_\t{a}^{(0)}\,\bar{a}_\mp\right)\delta \tilde{u} \\
				& + \sqrt{(1-\eta_\t{c})\kappa}\left(\delta \tilde{s}_\t{vac}^\pm +
					\frac{i\gamma}{2}\chi_\t{a}^{(0)}\, \delta \tilde{s}_\t{vac}^\mp\right) \\
				& + \sqrt{\eta_\t{c} \kappa}\left(\delta \tilde{s}_\t{in}^\pm + 
					\frac{i\gamma}{2}\chi_\t{a}^{(0)}\delta \tilde{s}_\t{in}^\mp\right)
	\end{aligned}
	\end{equation}
	\begin{equation}\label{eq:fluctuations_mechanics_fourier}
		\left(\chi_\t{m}(\Omega)^{-1}+\chi_\t{fb}(\Omega)^{-1}+\chi_\t{ba}(\Omega)^{-1}\right)\delta \tilde{u} = 
			\delta \tilde{f}_\t{th} + \delta\tilde{f}_\t{fb,th}+ \delta\tilde{f}_\t{ba,th}.
	\end{equation}
\end{subequations}
Here $\chi_\t{fb}$ and $\chi_\t{ba}$ are the modification to the intrinsic mechanical susceptibility due to 
feedback and dynamic back-action, respectively.  Likewise $f_\t{fb,th}$ and $f_\t{ba,th}$ represent effectively 
thermal components of the feedback and measurement back-action forces, respectively, adopting the notation from 
\secref{sec:theoryFeedback}. Before elaborating, we emphasize the following simplifications in the experimentally 
relevant `bad-cavity' limit, $\kappa \gg \Omega_\t{m}$, assuming a resonantly driven cavity ($\Delta=0$) 
and adopting the dissipative feedback strategy described in \secref{sec:practicalfeedback}:
\begin{equation}\label{eq:susceptibilitiesZero}
\begin{split}
	& \chi_\t{a}^{(0)}(\Omega)^{-1} \eqdef -i(\Omega+\Delta)+\frac{\kappa}{2} \approx \frac{\kappa}{2}\\
	& \chi_\t{a}^{(\gamma)}(\Omega)^{-1} \eqdef 
		\frac{\chi_\t{a}^{(0)}(\Omega)^{-1}}{\chi_\t{a}^{(0)}(\Omega)^{-2}+(\gamma/2)^2} 
		\approx \frac{\kappa}{2}\left(1+\frac{\gamma^2}{\kappa^2}\right) \\
	& \chi_\t{m}(\Omega)^{-1} := \Omega_\t{m}^2 -\Omega^2 -i\Omega \Gamma_\t{m}\\
	& \chi_\t{fb}(\Omega)^{-1}:= \Omega_\t{fb}^2(\Omega) -i\Omega \Gamma_\t{fb}(\Omega) \approx -i\Omega \Gamma_\t{m}(1+g_\t{fb})\\
	& \chi_\t{ba}(\Omega)^{-1} := \Omega_\t{ba}^2(\Omega) -i\Omega \Gamma_\t{ba}(\Omega) \approx 0.
\end{split}
\end{equation}

\subsection{Dynamic back-action}\label{sec:probeDBA}

When the cavity is driven away from resonance ($\Delta \neq 0$), correlations between the radiation pressure back-action force and the mechanical position give rise to a well known dynamic radiation pressure back-action force \cite{aspelmeyer2013}. In the high-Q ($\Omega_\t{m}\gg\Gamma_\t{m}$), bad-cavity ($\kappa \gg \Omega_\t{m}$) limit relevant to our experiment, dynamic back-action manifests as a displaced mechanical frequency (the optical spring effect) and passive cold-damping \cite{aspelmeyer2013}.  Accounting for cavity mode splitting, the optically-induced spring shift ($\Delta\Omega_\t{ba}$) and damping rate ($\Gamma_\t{ba}$) are given by:
\begin{subequations}\label{eq:probeDBA}
\begin{equation}\label{eq:probeDBA_spring}
	\Delta\Omega_\t{ba}\eqdef\Omega_\t{ba}(\Omega_\t{m})-\Omega_\t{m}\approx \frac{2g_0^2}{\kappa} \frac{4\eta_\t{c} P_\t{in}^+}{\kappa \hbar \omega_\t{c}}
		\sum_{j=\pm} \frac{(\kappa/2)^3 (\Delta + j\gamma/2)}{\left[(\Delta +j\gamma/2)^2 +(\kappa/2)^2\right]^2}
\end{equation}
\begin{equation}\label{eq:probeDBA_damping}
	\Gamma_\t{ba}(\Omega_\t{m})\approx \frac{\Omega_\t{m}}{4\kappa}\cdot\frac{2g_0^2}{\kappa} \frac{4\eta_\t{c} P_\t{in}^+}{\kappa \hbar \omega_\t{c}}
		 \sum_{j=\pm} \frac{\kappa^5(\Delta-j\gamma/2)}{\left[(\Delta+j\gamma/2)^2+(\kappa/2)^2\right]^3}.
\end{equation}
\end{subequations}
\eqref{eq:probeDBA_spring} is used in conjunction with \eqref{eq:transmission} to estimate $g_0$ in section \secref{sec:g0}. 
 Note that both terms vanish for resonant probing.

\subsection{Stochastic back-action}\label{sec:probeQBA}

When the cavity is driven on resonance ($\Delta = 0$), the thermal component of the radiation pressure back-action force takes the form
\begin{equation}\label{eq:fQBA}
\begin{split} 
	\delta \tilde{f}_\t{ba,th} = \frac{8g_0 \Omega_\t{m} }{\sqrt{\kappa}\left(1+\gamma^2/\kappa^2\right)}
		\left\{
		 \left(\sqrt{n_+}+\frac{\gamma}{\kappa}\sqrt{n_-}\right)\sqrt{\eta_\t{c}} \delta \tilde{q}_\t{in}^{+} 
		  + \left(\sqrt{n_+}+\frac{\gamma}{\kappa}\sqrt{n_-}\right)\sqrt{1-\eta_\t{c}} \delta \tilde{q}_\t{vac}^{+}  \right.\\ 
		\left. -\left(\frac{\gamma}{\kappa}\sqrt{n_+}-\sqrt{n_-}\right)\sqrt{\eta_\t{c}} \delta \tilde{p}_\t{in}^{-}
		-\left(\frac{\gamma}{\kappa}\sqrt{n_+}-\sqrt{n_-}\right)\sqrt{1-\eta_\t{c}} \delta \tilde{p}_\t{vac}^{-}
		\right\},
\end{split} 
\end{equation}
where $q(p)$ denote the amplitude(phase) quadrature of each field: $\delta s = \delta q + i \delta p$. In \eqref{eq:fQBA}, 
we have retained the explicit dependence on $n_{\pm}$ in order to emphasize their role in weighting the various noise components. 
We note that as a consequence of the scattering process, (amplitude)phase fluctuations entering the
(clockwise)counter-clockwise mode are converted to intensity fluctuations by two pathways. 

Assuming that the drive field is shot-noise limited in its amplitude quadrature ($\bar{S}_{qq}^\t{in} = \tfrac{1}{2}$) and that the cavity is otherwise
interacting with a zero temperature bath ($\bar{S}_{qq}^\t{vac} = \tfrac{1}{2}=\bar{S}_{pp}^\t{vac}$), we find that the effective thermal occupation of the remaining `quantum' stochastic back-action is given by
\begin{equation}\label{eq:nQBA}
	n_\t{ba} = C_0 \frac{1}{1+\gamma^2/\kappa^2}(n_+ + n_-) = C_0 n_+;
\end{equation}
here expressed in terms of the `single-photon cooperativity' parameter,
\begin{equation}
	C_0 \eqdef \frac{4g_0^2}{\kappa \Gamma_\t{m}}.
\end{equation}

\subsection{Measurement imprecision}\label{sec:probeImp}

The cavity transmission, $\delta \tilde{s}^{+}_\t{out} = \delta \tilde{s}^{+}_\t{in}-\sqrt{\eta_\t{c} \kappa}\,\delta
\tilde{a}_+$, at 
$\Delta = 0$ is given by,
\begin{equation}\label{eq:sout}
\begin{split}
	\delta \tilde{s}^{+}_\t{out} = -i\sqrt{\eta_\t{c}}\frac{2g_0\sqrt{n_+}}{\sqrt{\kappa}}\, 
			\left(\frac{1-\gamma^2/\kappa^2}{1+\gamma^2/\kappa^2}\right) \delta \tilde{u} +
			\left(1-\frac{2\eta_\t{c}}{1+\gamma^2/\kappa^2}\right)\, \delta \tilde{s}_\t{in}^{+} 
			& -i\frac{2\eta_\t{c} (\gamma/\kappa)}{1+\gamma^2/\kappa^2}\, \delta s_\t{in}^{-} \\
			&  -\frac{2\sqrt{\eta_\t{c} (1-\eta_\t{c})}}{1+\gamma^2/\kappa^2}\left(\delta \tilde{s}_\t{vac}^{+}
			+ i\frac{\gamma}{\kappa}\delta \tilde{s}_\t{vac}^{-}\right).
\end{split}
\end{equation}
As depicted in \fref{fig:theorySchematic}, the transmitted field is amplified in a balanced homodyne receiver with 
a coherent local oscillator (LO) $s_\t{lo}$. The fields transmitted at either ports of the homodyne beam-splitter are,
\begin{equation}
	\begin{pmatrix}
		\tilde{s}_1 \\ \tilde{s}_2
	\end{pmatrix} = \frac{1}{\sqrt{2}}
	\begin{pmatrix}
		1 & i \\ i & 1
	\end{pmatrix}
	\begin{pmatrix}
		s_\t{lo} \\ \delta \tilde{s}_\t{out}^+
	\end{pmatrix};
\end{equation}
the optical intensities detected by independent identical photodetectors are, 
$\delta \tilde{I}_i = \tilde{s}_i^\dagger \tilde{s}_i$ ($i=1,2$).
The operator corresponding to the resulting subtracted homodyne intensity is, 
\begin{equation}
	\delta \tilde{I}_\t{hom} = \delta \tilde{I}_1 -\delta \tilde{I}_2 = 2\vert s_\t{lo}\vert \left(\delta \tilde{p}_\t{out}^+ \cos \theta_\t{lo}
			- \delta \tilde{q}_\t{out}^+ \sin \theta_\t{lo}\right),
\end{equation}
where $\vert s_\t{lo}\vert$ is the amplitude of the large coherent LO field, and $\theta_\t{lo}$ the
relative mean phase between the LO and the cavity transmission. The path length of the LO arm is electronically
locked to maintain $\theta_\t{lo} \approx 0$, so that the homodyne signal picks out the phase quadrature of the cavity transmission.
For photodetectors with gain $G_\t{d}$ (A/W) and quantum effeciency $\eta_\t{d}$, the resulting shot-noise-normalized spectrum of 
photocurrent fluctuations is given by \cite{Carm87}:
\begin{equation}
\begin{split} 
	S_{i}^\t{hom}(\Omega) &= G_\t{d}^2 \eta_\t{d} \left(1 + \eta_\t{d}
			\frac{\langle :\delta \tilde{I}_\t{hom}(\Omega)\, \delta \tilde{I}_\t{hom}(-\Omega):\rangle }{\langle I_\t{hom}\rangle} 
		\right) \\
		&= G_\t{d}^2 \eta_\t{d} \left(1 + \eta_\t{d}\eta_c\frac{16g_0^2 n_+}{\kappa}
			\left(\frac{1-\gamma^2/\kappa^2}{1+\gamma^2/\kappa^2}\right)^2 S_u (\Omega) \right).
\end{split}
\end{equation}
Using \eqref{eq:Simpdef} and $u\equiv x/x_\t{zp}$, the shot noise floor of the homodyne photocurrent spectrum can be expressed as an equivalent thermal bath occupation,
\begin{equation}\label{eq:nImp}
	n_\t{imp} = \left(\frac{1}{16\eta_\t{c}\eta_\t{d} C_0 n_+}\right)
		\left(\frac{1+\gamma^2/\kappa^2}{1-\gamma^2/\kappa^2} \right)^2.
\end{equation}
Note that mode splitting causes the optical susceptibility (\eqref{eq:fluctuations_probe}) to flatten near resonance, leading to divergence of \eqref{eq:nImp} when $\gamma=\kappa$.

\subsection{The uncertainty principle and the standard quantum limit}\label{sec:SQL}

\eqref{eq:nQBA} and \eqref{eq:nImp} imply that a cavity-optomechanical position measurement is bound by the imprecision-back-action product,
\begin{equation}\label{eq:uncertainty}
4\sqrt{n_\t{imp}n_\t{ba}}=\frac{1}{\hbar}\sqrt{S_x^\t{imp}S_F^\t{ba}}\ge 1.
\end{equation}
Using \eqref{eq:SFFdef} and \eqref{eq:Szp}, we identify
\begin{subequations}\begin{align}
S_x^\t{imp}&= n_\t{imp}\cdot 2S_x^\t{zp}\\
S_F^\t{ba}&= n_\t{ba}\cdot m^2|\chi(\Omega_\t{m})|^{-2}\cdot 2S_x^\t{zp}=4\hbar n_\t{ba}\Omega_\t{m}\Gamma_\t{m}m
\end{align}
\end{subequations}
as, respectively, the shot-noise limited imprecision of the homodyne measurement, \eqref{eq:nImp}, (referred from the photocurrent to the mechanical position) and its associated stochastic back-action, arising from radiation pressure shot noise, \eqref{eq:backaction}.

This product places a limit on the apparent motion of the oscillator.  Namely, in the absence of feedback, \eqref{eq:SxSxmeas} gives    

\begin{equation}\label{eq:SQL}
\frac{S_y(\Omega_\t{m})}{2S_x^\t{imp}}=n_\t{th}+n_\t{ba}+n_\t{imp}+\frac{1}{2}\ge n_\t{th}+1\ge1.
\end{equation}
The limiting case in \eqref{eq:SQL} occurs for $n_\t{th}=0$ and $n_\t{ba}=n_\t{imp}=\tfrac{1}{4}$.  This corresponds to an oscillator in contact with a zero-temperature ambient thermal bath, measured with an imprecision of $S_x^\t{zp}/2$, and exhibiting, due to stochastic back-action, a physical displacement of $S_x^\t{zp}/2$ on top of its zero-point displacement, of magnitude $S_x^\t{zp}$.  

A more general treatment \cite{Clerk10} reveals the RHS of \eqref{eq:uncertainty} and the RHS of \eqref{eq:SQL} to coincide with the Heisenberg uncertainty principle and the standard quantum limit for a weak continuous (linear) position measurement, respectively.


\vfill \pagebreak
\section{Experimental details}\label{sec:Experimental}

\subsection{Sample design and fabrication}\label{sec:Fabrication}


Our optomechanical system consists of a doubly clamped Si$_3$N$_4$ beam of length $L\approx68$ $\mu$m, width
$w\approx400$ nm, and thickness $t\approx70$ nm placed $z\sim50$ nm above the surface of a wedged silica microdisk
\cite{lee_chemically_2012} with radius $R_\t{d}\approx 15$ $\mu$m, thickness $t_\t{d}\approx0.65$ $\mu$m, and wedge angle
$\theta_\t{d}\approx 30$ degress.  Beam and microdisk are monolithically integrated on a Si microchip
The dimensions of the system were chosen with the aim of maximizing single-photon cooperativity
$C_{0}=4G^2x_{\t{zp}}^2/\kappa\Gamma_{\t{m}}$ for the fundamental out-of-plane mode of the nanomechanical beam. 
Towards this end, a crucial consideration is the co-locolization of the mechanical and optical mode volumes.  Finite
element modeling (COMSOL 4.3) was used to compute the field distribution of the whispering gallery optical mode and its
overlap with the mechanical mode; the vertical gradient of this overlap integral is proportional to the frequency
pulling factor $G$ \cite{anetsberger2009}.  As a rule of thumb, $G$ is increased by centering the lateral
position of the beam ($x$) within the evanescent optical mode and by minimizing the vertical separation ($z$) between
the beam and surface of the disk (Fig. 1 of main text).  Optimal values of $\{t,w\}$ are determined by maximizing $G
x_{zp}$ while fixing all other dimensions, in this case leveraging the trade-off between increasing mode overlap and
mechanical mass.  $G$ may also be increased by decreasing $t_\t{d}$ and $R_\t{d}$; 
this enhancement however comes at the cost of increased optical losses, including waveguide coupling to the beam. The
sample under study was thus chosen from an experimental sweep of $C_0$ versus $\{R_\t{d},w,x\}$.  The chosen value of $L$
inherits from fabrication constraints as well as an effort to localize the frequency of the fundamental out-of-place
mode in a region of low extraneous noise (Fig. 1D of main text).

\begin{figure*}[h!] 
	\vspace{-20pt}
	\centering
	\includegraphics[width=0.35\linewidth]{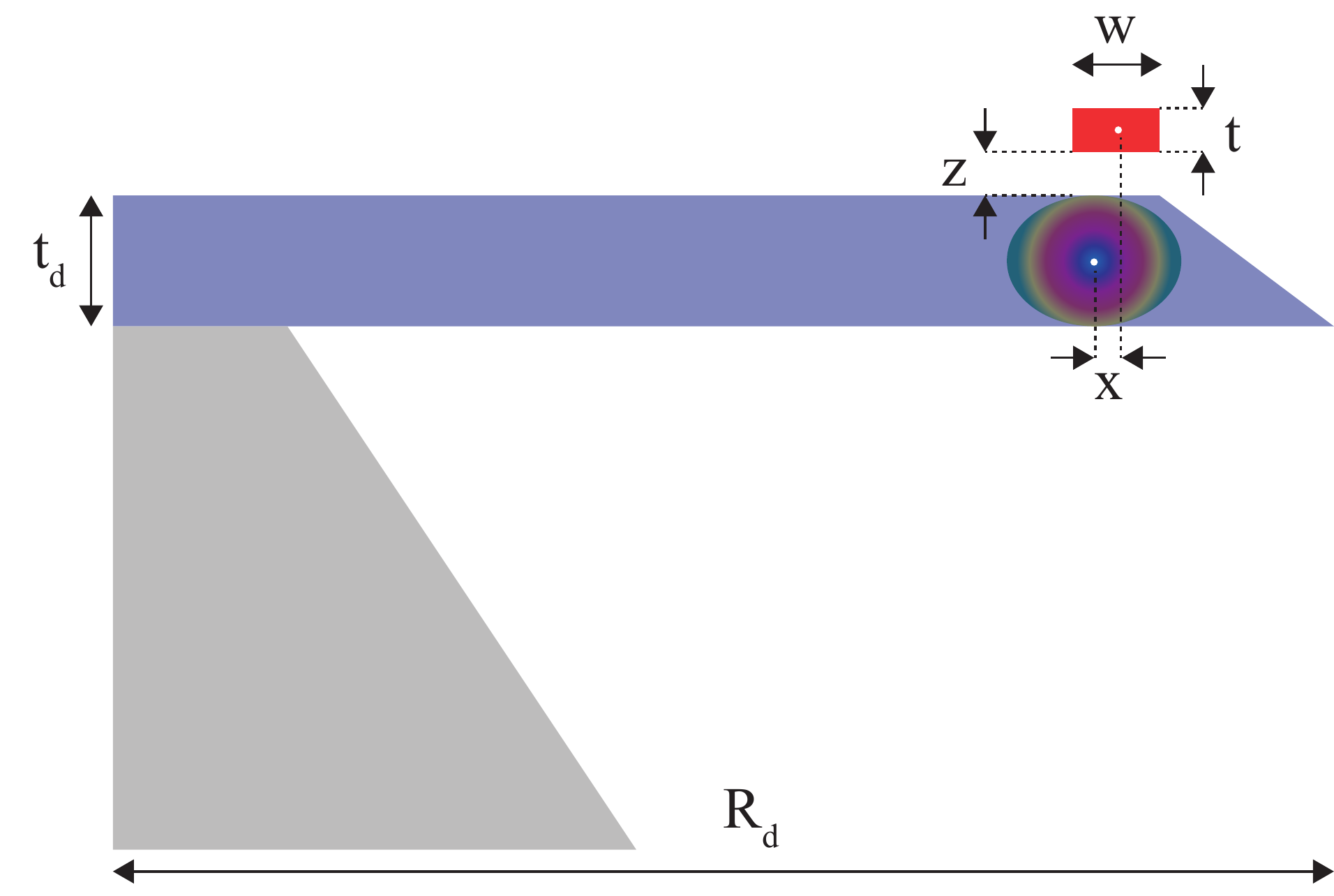}
	\caption{\label{fig:SI:geometry} Microdisk-nanobeam geometry}
\end{figure*}

The fabrication process begins by dry oxidation of a float zone Si wafer, in order to grow a high purity SiO$_2$ film.
The microdisk and nanobeam support pads are defined by photolithography and transferred to the SiO$_2$ by wet etching in
buffered hydrofluoric acid. A second photolithography and wet etching step is used to selectively thin the microdisk
only, which defines the gap between the nanobeam and microdisk. An etch-stop layer of Al$_2$O$_3$ is deposited by atomic
layer deposition, in order to protect the microdisk later in the process, when the nanobeam is etched. Next a poly-Si
layer is deposited and planarized by chemical mechanical polishing. The planarization is stopped when the nanobeam
support pads are exposed, but a thin sacrificial layer of poly-Si still remains above the microdisk. Afterwards, a high
stress Si$_3$N$_4$ film is deposited by low pressure chemical vapor deposition. A series of steps are carried out to
expose alignment marks defined in the SiO$_2$ layer. The nanobeam pattern is defined and precisely positioned, using the
alignment marks, by electron beam lithography. The Si$_3$N$_4$ is etched with SF$_6$ chemistry, using an inductively
coupled plasma etcher. Finally the microdisk and nanobeam are undercut by etching in a potassium hydroxide solution,
followed by critical point drying.

\subsection{Cryogenic Operation}\label{sec:cryostat}

The sample is embedded in a $^3$He buffer gas cryostat (Oxford Instruments HelioxTL).  As detailed in Riviere
\emph{et. al.} \cite{riviere_evanescent_2013}, laser light is coupled to the microdisk by means of a straight tapered
optical fiber affixed to the cryostat probe head.  To position the microdisk relative to the optical fiber, the sample
chip is mounted on a 2-axis Attocube nanopositioner (ANPx50/LT). An important practical consequence of the near-field
coupling architecture is that it enables us to place the tapered fiber in physical contact with the microdisk without
influencing the quality factor of the nanobeam.  We operate in this ``contact mode'' in order to suppress fluctuations
in the coupling strength $\kappa_{ex}$ due low frequency cryostat vibrations, as well as drift due to temperature
change.  As shown in Fig. \ref{fig:SI:transmission_v_kappa}, changing the position of the contact point allows access to
a wide range of coupling strengths, including nearly ideal \cite{spillane2003} critical coupling. 

We regulate the pressure and temperature of our cryostat in order to address different experimental challenges. 
Measurements which require independent knowledge of the sample temperature (e.g. optomechanical coupling,
\secref{sec:g0}) are performed using a large buffer gas pressure of $\sim 100$ mbar in order to ensure good thermalization
with the sample holder, whose temperature is monitored using a calibrated Cernox sensor.  Measurements requiring high
mechanical quality factor are conducted with the buffer gas evacuated to a pressure of $<10^{-3}$ mbar.  We have
verified that the sample remains thermalized with the sample holder at temperatures as low as 4 K 
(\fref{fig:SI:thermalization}), by monitoring the thermomechanical noise of multiple beam modes using a weak sensor field
(Fig. \ref{fig:SI:thermalization}).  Below 4 K, a dramatic rise in temperature is observed for all modes (inset to
\fref{fig:SI:thermalization}).  This temperature rise scales linearly with optical power, and suggests an increase in
susceptibility to absorption heating.  We conjecture that this effect is due to a rapid drop in thermal conductivity
consistent with the universal behavior of amorphous glass (in this case Si$_3$N$_4$) at temperatures below $\sim10$K 
\cite{pohl2002low}.  To avoid this strong effect, we operate in temperature ranging from
4-5 K for all of the reported experiments.

\begin{figure}[h!]
	\centering
	\includegraphics[scale=0.6]{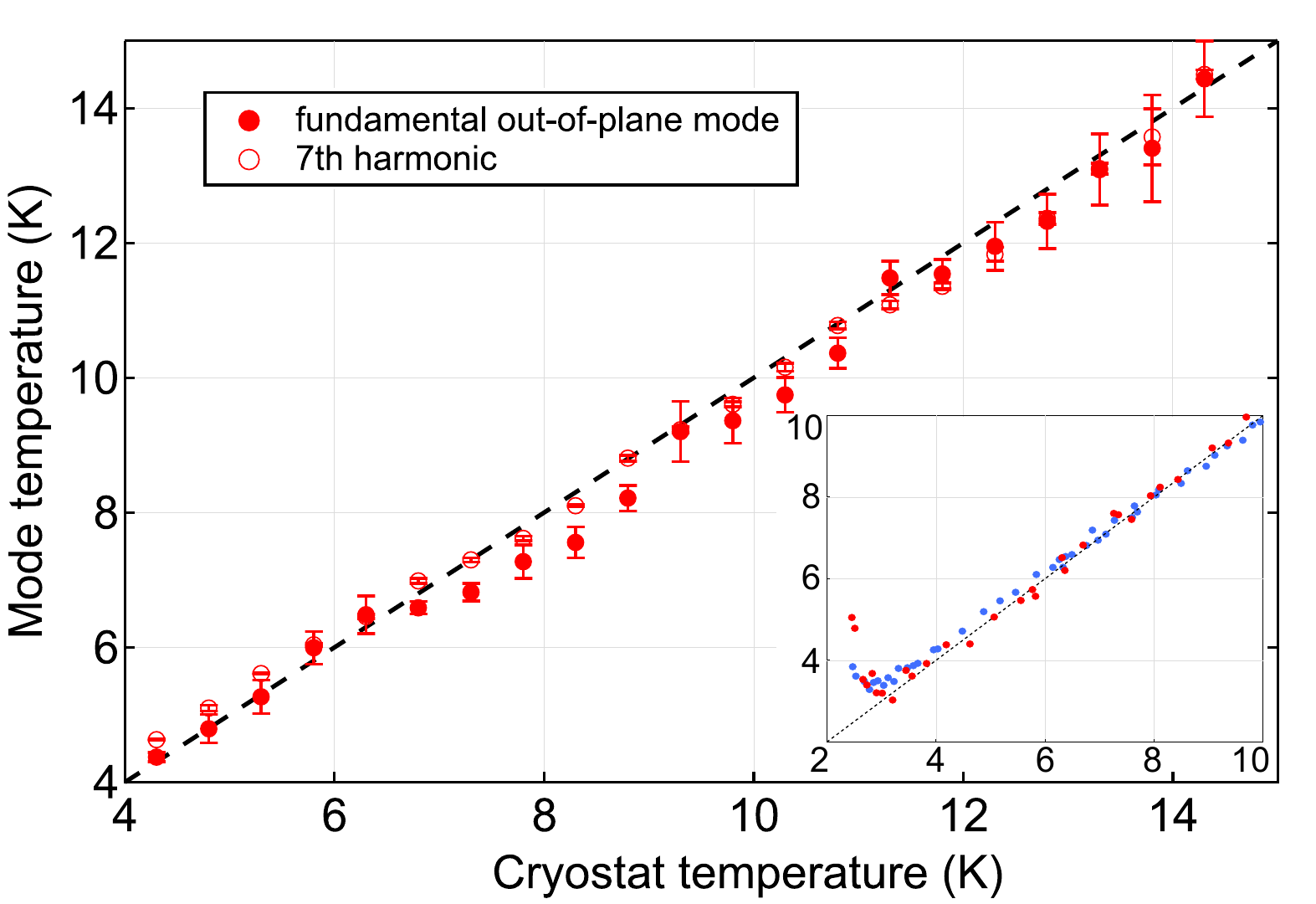}
	\caption{\label{fig:SI:thermalization} Mode temperature vs. cryostat temperature.}
\end{figure}

\subsection{Experimental setup}

\begin{figure}[t]
	\centering
	\includegraphics[width=0.9\linewidth]{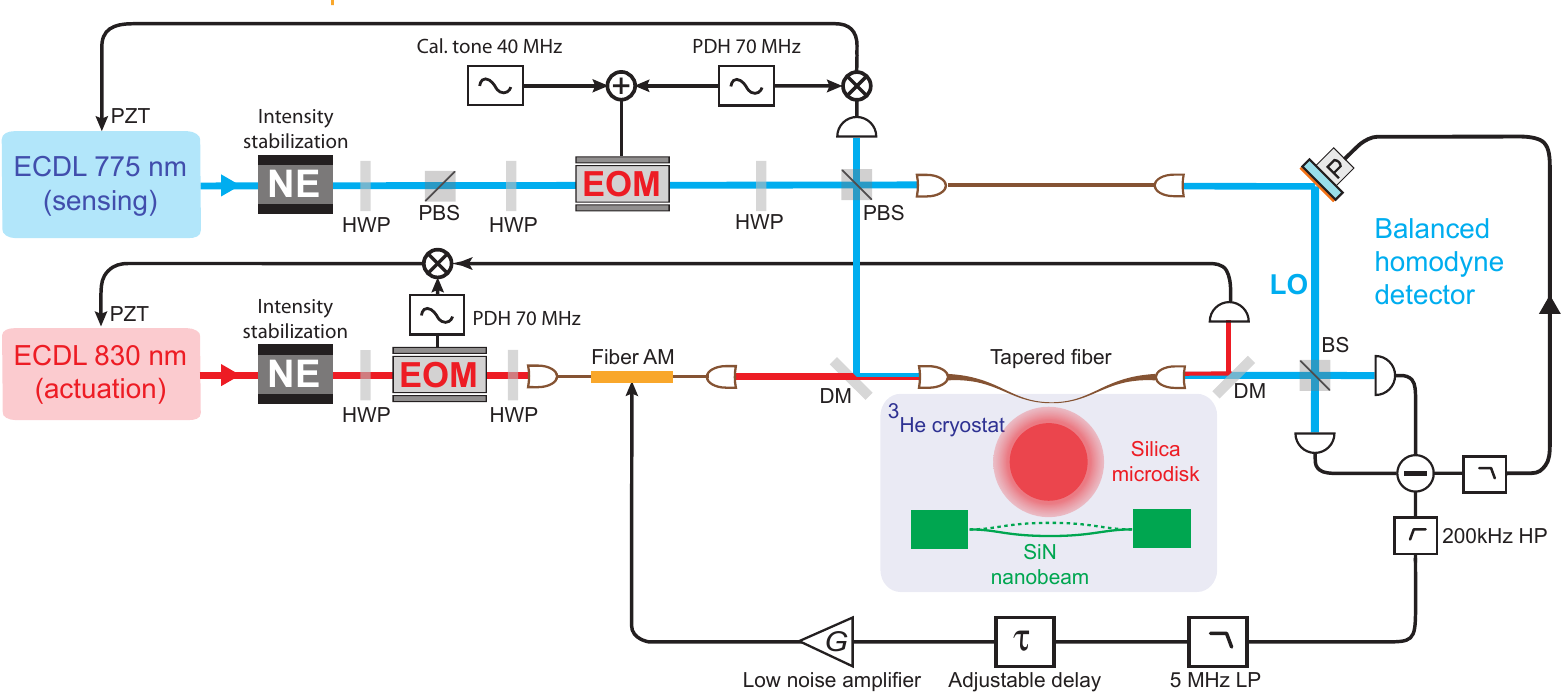}
	\caption{\label{fig:SI:setup} Schematic of experimental setup}
\end{figure}

A schematic of the experiment setup is shown in \fref{fig:SI:setup}.  At its heart is the cryogenic, taper-coupled
microdisk (\secref{sec:cryostat}).  The taper is spliced into a $\sim 9$ meter, single mode optical fiber (780HP),
penetrating the cryostat probe through a pair of teflon vacuum feedthroughs. 
Optical fields coupled to the fiber are supplied by two intensity-stabilized
(Thorlabs NE LCC3112H) external cavity diode lasers (ECDL, New Focus Velocity 6312 and 6316) operating at
$\lambda\approx$ 775 nm and 850 nm: the `sensor' and `feedback' laser, respectively. Each laser is phase modulated using
a broadband EOM (PM, New Focus 4002).  Phase modulation at 61(42) MHz is used to generate a PDH error signal with which
to stabilize the sensor(feedback) laser frequency to the sensor(feedback) cavity mode.  For the sensor field, a second
phase modulation tone at $\Omega_{\t{cal}}/2\pi = 40$ MHz is used to calibrate the homodyne measurement (Sec.
\ref{sec:g0}).  Directly before(after) the signal fiber, the sensor and feedback fields are combined(split) using a
dichroic mirror (DM).  The sensor(feedback) PDH error signal is derived from a weak pickoff of the
reflected(transmitted) cavity field. 

The homodyne detector is formed by incorporating the cryogenic signal fiber into one arm of a carefully
length-matched Mach-Zehnder interferometer.  The local oscillator (LO) arm of the interferometer is derived from the
sensor field using a beamsplitter located after the broadband EOM.  Cancelling the signal generated by common-mode phase
modulation on a single port of the balanced photodetector (FEMTO HCA-S) enables us to match the length of the signal and
LO arm to within 1 mm, practically eliminating contamination of the measurement by laser phase noise.  Subsequent power
balancing of the two detection ports achieves a common-mode rejection of residual amplitude modulation --- mainly
arising due to optical fiber etalons --- by $\sim30$ dB.  The balanced photodetector, based on a matched pair of Si PIN
photodiodes, features a low gain, DC-coupled transimpedance amplifier and a high gain ($5\cdot 10^4$ V/A) AC-coupled
transimpedance amplifer with a low NEP of $\sim 10$ pW/$\sqrt{\t{Hz}}$ at 5 MHz.  We use a LO power of 4 mW to
achieve a shot noise to detector noise ratio of $\gtrsim 6$ dB at Fourier frequencies near $\Omega_{\t{m}}/2\pi=4.32$ MHz. 
The DC photosignal is used to stabilize the path length of the interferometer by feedback to a piezo-actuated mirror
in the LO path. 

For characterization of measurement imprecision, the AC homodyne photosignal is sent directly to the spectrum analyzer
(SA, Tektronix RSA5106A).  For feedback cooling, the photosignal is split on a 20 dB directional coupler (Minicircuits
ZFDC-20-3+).  The weak port is sent to the SA.  The strong port is directed to a low $V_\pi(\approx 5$ V$)$ fiber
intensity modulator (EOSPACE) in the feedback beam path.  For the signal(feedback) power used in the reported cooling
experiment, $5.5(0.1)$ $\mu$W, it was necessary to further amplify the photosignal in order to achieve the largest
reported damping rates.  A low noise voltage amplifier (Minicircuits ZFL-500LN) was thus placed after the directional
coupler, followed by a voltage-controlled RF attenuator (Minicircuits ZX73-2500-S+), used to tune the feedback gain.  In
order to suppress feedback to higher-order beam modes, the photosignal was also passed through a 5 MHz low-pass filter
(Minicircuits BLP-5+).  The remaining electronic path length was manually fine tuned, by minimizing the feedback spring
effect, to achieve a total feedback delay of $3\pi/2\Omega_{\t{m}}\approx$ 175 ns.

\subsection{Calibration of optomechanical coupling rate $g_0$}\label{sec:g0}


We determine the zero-point optomechanical coupling rate $g_0$ of our system by calibrating the transduction factor
$G_{V\omega}$ connecting thermomechanical cavity frequency noise, $S_\omega(\Omega) \approx
8g_0^2n_{\t{th}}/\Gamma_{\t{m}}\cdot |\chi_{\t{m}}(\Omega)/\chi_{\t{m}}(\Omega_{\t{m}})|^2$ with
the measured homodyne photocurrent noise, $S_V(\Omega) = |G_{V\omega}(\Omega)|^2 S_\omega(\Omega)$ (here photocurrent has
been referred to the voltage $V$ measured at the output of the photodetector transimpedance amplifier).  Following the
method detailed in \cite{gorodetsky_determination_2010}, we take advantage of the fact that the cavity transduces laser
frequency fluctuations and cavity frequency fluctuations in the same way.  To calibrate $G_{V\omega}$, we phase-modulate
the sensor field at frequency $\Omega_{\t{cal}}$ with a known modulation depth $\beta$; this produces a reference
tone of magnitude $S_{V}^\t{cal}(\Omega) = \tfrac{1}{2}\Omega_\t{cal}^2\beta^2\delta(\Omega-\Omega_{\t{cal}})|G_{V\omega}(\Omega)|^2$.
Comparing the integrated area beneath the reference tone, $\avg{V^2}_\t{cal} = \tfrac{1}{2}\Omega_\t{cal}^2 \beta^2
\vert G_{V\omega}(\Omega_\t{cal}) \vert^2$, and the thermomechanical noise peak,
$\langle V^2\rangle_{\t{m}}=2g_0^2n_{\t{th}}|G_{V\omega}(\Omega_{\t{m}})|^2$, gives,
\begin{equation}
	g_0 = \frac{\beta\Omega_{\t{cal}}}{2}\sqrt{\frac{1}{n_{\t{th}}}
		\frac{\langle V^2\rangle_{\t{m}}}{\langle V^2\rangle_{\t{cal}}}}\,
		\left\vert \frac{G_{V\omega}(\Omega_{\t{cal}})}{G_{V\omega}(\Omega_{\t{m}})}\right \vert.
\label{eq:g0}
\end{equation}

An example of a $g_0$ measurement is shown in \fref{fig:SI:g0}. For this measurement, a buffer gas pressure of
$\sim$100 mbar was used to ensure good thermalization of the sample to probe head at $T \approx 3.3$ K ($n_{\t{th}}
\approx 1.6\cdot10^4$).  The resulting mechanical gas damping rate, $\Gamma_{\t{m}}\approx 2\pi\cdot 64$ kHz, also
allows us to ignore dynamic back-action effects for the moderate sensor power used, $P_{\t{in}}^+\approx 1$ $\mu$W. 
From separately determined $\beta\approx0.057$ (inferred from a heterodyne beat measurement), $\Omega_{\t{cal}}=
2\pi\cdot40$ MHz (the value of $\Omega_{\t{cal}}\gg\Omega_{\t{m}}$ was chosen in order to reduce residual
amplitude modulation),  and $|G_{V\omega}(\Omega_{c})|/|G_{V\omega}(\Omega_{\t{m}})|\approx0.98$, we infer $g_0
\approx 2\pi\cdot21$ kHz.

As an independent measure of $g_0$, we red-detune the sensor field and compare the resulting shift of the mechanical
frequency to a standard model for radiation pressure dynamical back-action \eqref{eq:probeDBA}.  For this measurement, the mechanical
damping rate was reduced by evacuating the buffer gas pressure to $<10^{-3}$ mbar (Sec. \ref{sec:damping rate}).  In
\fref{fig:SI:g0}, the observed spring shift $\Delta\Omega_{\t{ba}}$ is plotted versus cavity transmission
for an input power of $P_{\t{in}}^+\approx 1\, \mathrm{\mu W}$ and a measured cavity linewidth of $\kappa \approx 2\pi\cdot 1070$
MHz.  Incorporating the effect of measured cavity mode splitting (\secref{sec:cavity mode splitting}) into the model,
the measured spring shift is consistent with $g_0\approx 2\pi\cdot19$ kHz. The seperately determined values of 19 kHz
and 21 kHz (Fig \ref{fig:SI:g0}) are used to set error bars on estimates of $C_0$ and $n_\t{imp}$ in the reported
experiments (Table \ref{tab:Uncertainties}).

\begin{figure}[t]\label{fig:SI:g0}
	\centering
	\includegraphics[width=1.0\linewidth]{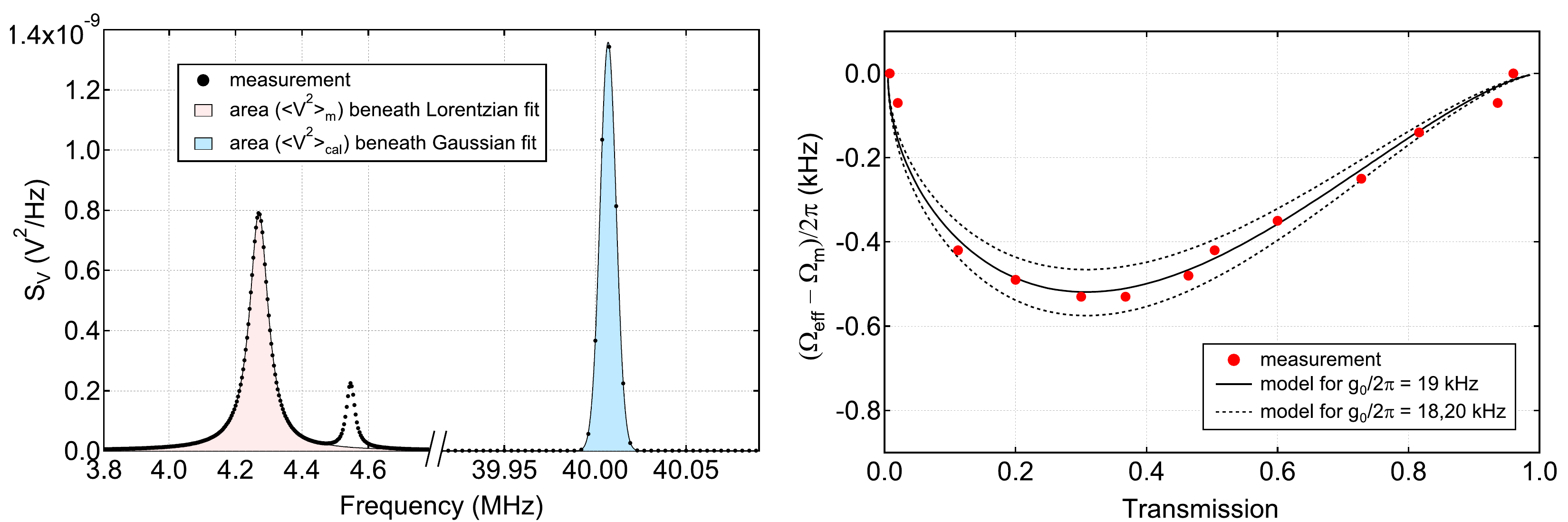}
	\caption{\label{fig:SI:setup} Calibration of $g_0$ by two methods.  Left: Using thermal noise and calibrated RF frequency modulation tone \cite{gorodetsky_determination_2010}.  Right: Using a model for the optical spring shift.}
\end{figure}

\subsection{Mechanical damping rate}\label{sec:damping rate}

To determine the intrinsic mechanical damping rate, $\Gamma_{\t{m}}$, it is necessary to minimize extraneous sources
of physical and apparent damping; these include gas pressure, radiation pressure and bolometric back-action, and slow
thermal drift of the mechanical frequency. To mitigate the former, we conduct experiments with the buffer gas in our
cryostat evacuated to a level $<10^{-3}$  mbar. We verify that the oscillator still thermalizes with the
sample holder to a temperature as low as 4 K (\fref{fig:SI:damping_rate}).  To mitigate back-action and drift, we
extract $\Gamma_{\t{m}}$ from an impulse-response measurement conducted with a weak probe (feedback) beam power of 
$<$ 50 nW. 

\begin{figure}[b]
	\centering
	\includegraphics[scale=0.6]{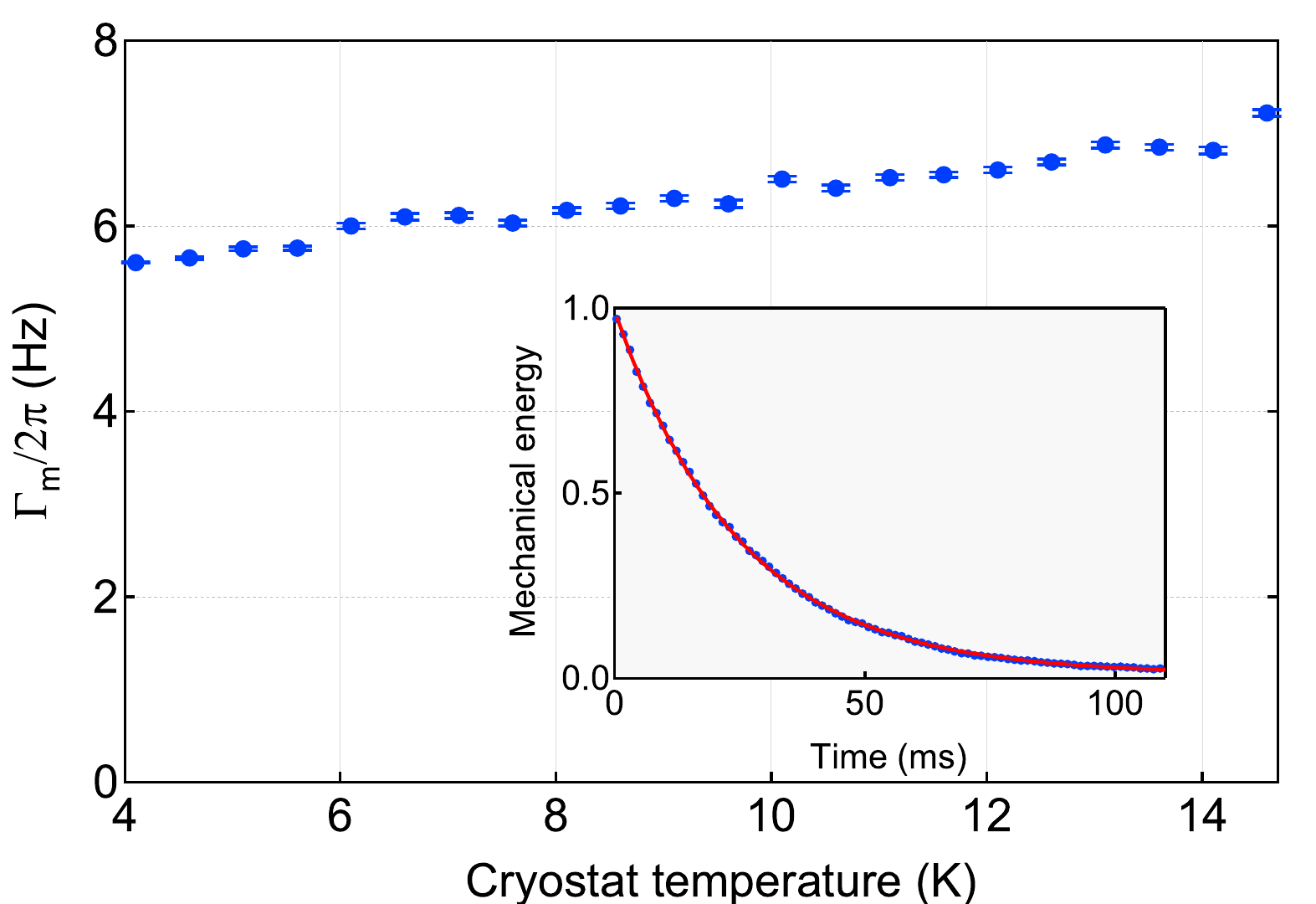}
	\caption{\label{fig:SI:damping_rate} Mechanical damping rate versus temperature. See Section \ref{sec:damping rate}. 
	Inset: ringdown example data (blue) and fit (red). }
\end{figure}

The step-response measurement is conducted as follows: 
the mechanical oscillator is driven with radiation pressure
by intensity modulating the feedback field at a frequency $\Omega_\t{d}\approx\Omega_{\t{m}}$.  
An RF switch is then used
to rapidly shutter off the modulation.  The subsequent exponential (ringdown) decay of the mechanical energy, with
e-folding time $\tau=2\pi/\Gamma_{\t{m}}$, is observed by demodulating the homodyne photocurrent at $\Omega_\t{d}$ with a
demondulation bandwidth of $B \gg \Gamma_{\t{m}}$.  An average of 100 such ringdowns in shown in the inset of
\fref{fig:SI:damping_rate}.  To record $\Gamma_{\t{m}}$ vs temperature in this figure, it was necessary to track the
frequency of the oscillator over a (temperature-induced) drift range of $\sim10$ kHz.  This was accomplished by incorporating the modulation
and demodulations signal into a phase-locked-loop, using a digital lock-in amplifier (Zurich Instruments UHFLI).  

\subsection{Mode splitting of probe cavity}\label{sec:cavity mode splitting}

To accurately estimate intracavity photon number, $n_\t{c}$, measurement imprecision, $n_\t{imp}$, and stochastic measurement
back-action, $n_\t{ba}$, it is necessary to account for cross-coupling between optical cavity modes. In a whispering
gallery microresonator, Rayleigh scattering from surface defects leads to coupling of otherwise degenerate clockwise
(CW) and counter-clockwise (CCW) propagating modes \cite{kippenberg2002} at a rate $\gamma$ (\eqref{eq:OMeqofmotion_optics}). 
Since only the clock-wise mode (by convention) is driven by the field from the optical taper, this leads to an effective 
reduction of photon collection efficiency by a factor $(1+\gamma^2/\kappa^2)^2$ (\eqref{eq:steadyState}).  
At $\Delta = 0$, coherence between the CW and CCW fields leads to a further decrease in homodyne readout sensitivity by a factor
$(1+\gamma^2/\kappa^2)^2/(1-\gamma^2/\kappa^2)^2$ (\eqref{eq:nImp}) and the simplified form for $n_\t{ba}$ given in \eqref{eq:nQBA}.  

As a confirmation of the coupled-mode model, we have characterized the steady state cavity transmission (\eqref{eq:transmission})
as a function of external coupling strength, $\kappa_{\t{ex}}$ (accessed by changing the taper's contact point on the
microdisk's surface). As shown in Fig. \ref{fig:SI:transmission_v_kappa}, the relationship of resonant transmission and
total decay rate $\kappa = \kappa_0+\kappa_{\t{ex}}$ shows good consistency with the model for an intrinsic decay
rate of $\kappa_0=2\pi\cdot440$ MHz and mode splitting $\gamma = 2\pi\cdot360$ MHz.  These values are used to analyze
data presented in Figures 1-3 of the main text.

\begin{figure}[h!]
	\centering
	\includegraphics[scale=0.6]{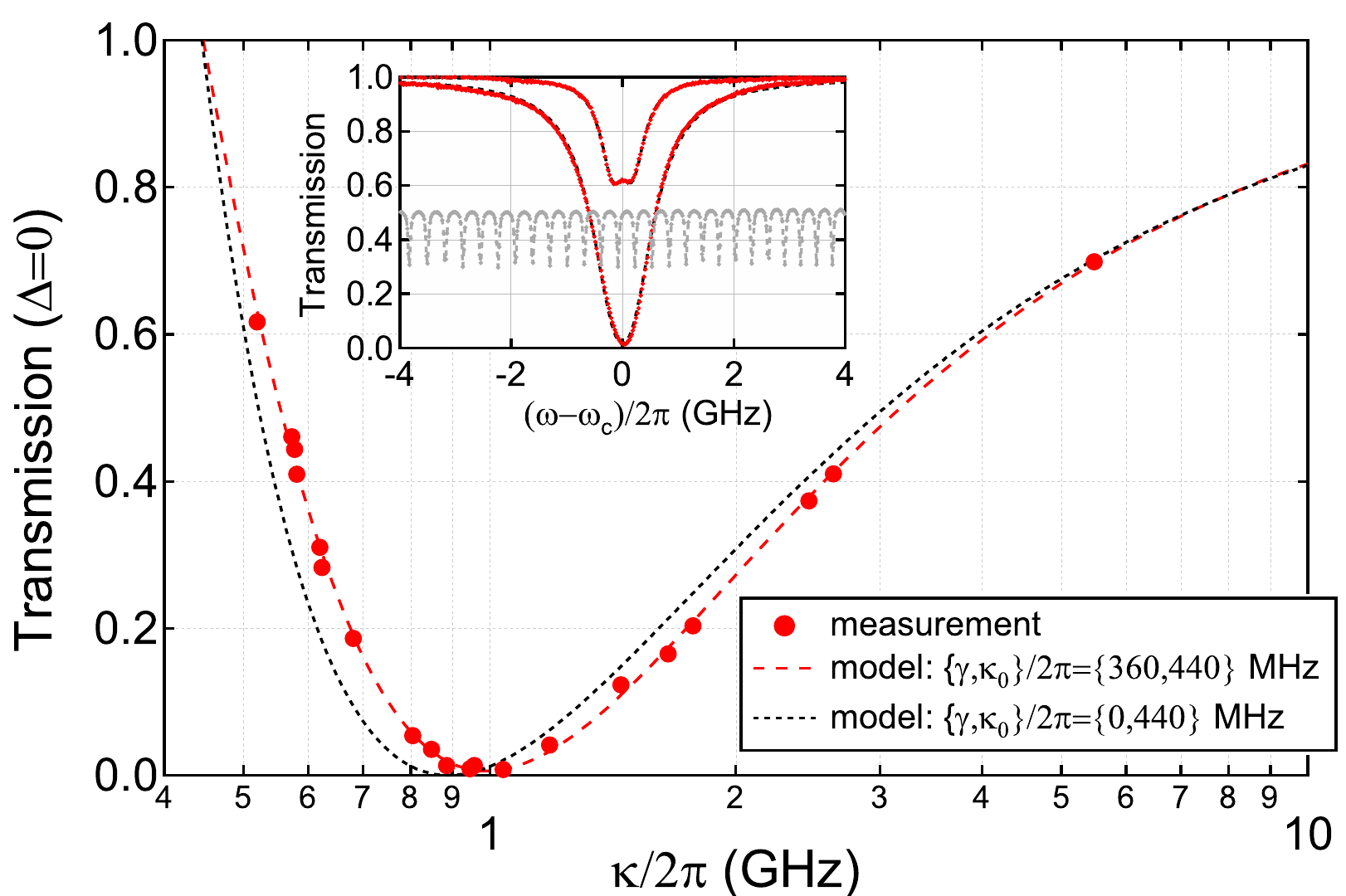}
	\caption{\label{fig:SI:transmission_v_kappa} Taper coupling ideality probed by cavity transmission. Inset: examples of
		cavity transmission (red) calibrated by using a pre-calibrated ``comb'' from a fiber loop cavity (gray).}
\end{figure}

\section{Summary of experimental values and systematic uncertainty}\label{sec:error}

Values used to determine experimental data points and their uncertainty in Figures
2-3 of the main text are summarized in Table \ref{tab:Uncertainties}.  We restrict our attention to sources of systematic
uncertainty, as these were found to dominate over statistical uncertainty (e.g. for least-squares fitting of
thermomechanical spectra).   
Uncertainties in $\{n_{\t{imp}}$, $n_{\t{tot}}\}$ (vertical axis of Fig. 2 of main text) are primarily due to
uncertainty in the value  the cryostat temperature.  Uncertainties in $\{n_{\t{eff}},n_{\t{fb}}\}$ (vertical axis
of Fig. 3 of main text) are primarily due to uncertainty in both the cryostat temperature and the magnitude of extraneous
back-action heating. Horizontal error bars in Figure 2 of the main text are primarily due to uncertainty in magnitude of
the vacuum optomechanical coupling rate. 

We highlight two sources of systematic uncertainty in the mechanical mode temperature: (1) discrepancy between the
cryostat Cernox sensor reading and the temperature at the location of the sample, and (2) heating due to extraneous
back-action. 
With regards to (1): two independently calibrated sensors placed in different locations on the sample holder read values
$T = 4.1$ K and $4.6$ K.  We take this to be the systematic uncertainty. With regards to (2): Extraneous back-action
heating is recorded versus optical power in Fig. 2 of the main text.  A similar measurement was made prior to the
cooling experiment shown in Fig. 3.  In the latter case, we observed approximately $0.4$ K of measurement back-action
heating for the $P_\t{in}^+\approx5.5$ $\mu$W sensor power used.  This is a factor of $\sim$2 smaller than shown in Fig. 2.  The
discrepancy is partly attributable to the use of a larger optical linewidth, $\kappa\approx 2\pi\cdot 1.85$ GHz, in the
feedback experiment.  As a conservative estimate, we assume a back-action heating of $(0.4,1)$ K for figure 3.

An estimate for the systematic uncertainty in $\Gamma_\t{m}$ is made by taking the extreme values $2\pi\cdot(5.6,5.8)$ Hz in
the range $T = (4.1, 4.6)$ K in Fig. \ref{fig:SI:damping_rate}. We note that in Fig. 2 of the main text, extraneous
back-action leads to a rise in effective bath temperature by as much as $\sim 12$ K, corresponding to $\Gamma_\t{m} \approx 2\pi \cdot 6.6$ Hz. As we do not have a model for the influence of such localized heating on $\Gamma_\t{m}$, we have chosen to omit this possible source of uncertainty from the treatment in the main text.

\begin{table*}[t]
\begin{tabular}{|p{1.1cm}|p{2.8cm}|p{4.2cm}|p{8cm}|}
\hline 
Symbol & Expression used & Value & Measurement method and source of systematic uncertainty \tabularnewline
\hline 
\hline 
$\lambda$& & $775$ nm & Wavelength meter.
\tabularnewline
\hline 
$\kappa$& & $2\pi\cdot910$ MHz (Fig. 2)\newline $2\pi\cdot1850$ MHz (Fig. 3)& Fit to transmission versus detuning (Sec. \ref{sec:cavity mode splitting}).
\tabularnewline
\hline 
$\gamma$ & & $2\pi\cdot360$ MHz & Fit to transmission versus detuning (Sec. \ref{sec:cavity mode splitting}).
\tabularnewline
\hline 
$\kappa_0$ & & $2\pi\cdot440$ MHz & Fit to resonant transmission versus $\kappa$ (Sec. \ref{sec:cavity mode splitting}).
\tabularnewline
\hline 
$T$ & & $(4.1,4.6)$ K & Reading from two independent cryostat thermistors (Sec. \ref{sec:error}).
\tabularnewline
\hline 
$\Omega_{\t{m}}$ & & $2\pi\cdot4.32$ MHz& Fit to thermomechanical noise peak.
\tabularnewline
\hline 
$\Gamma_\t{m}$ & & $2\pi\cdot(5.6,5.7)$ Hz & Mechanical ringdown, uncertainty due to temperature dependence in range  $T=(4.1,4.6)$ K.  
\tabularnewline
\hline 
$n_{\t{th}}$ & $k_B T/\hbar\Omega_{\t{m}}$ & $(2.0,2.2)\cdot10^4$ & Inferred from $\{\Omega_{\t{m}},T\}$. Uncertainty taken from $T$.
\tabularnewline
\hline 
$g_{0}$ & & $2\pi\cdot(19,21)$ kHz & Two independent calibration methods (Sec. \ref{sec:g0})
\tabularnewline
\hline 
$C_{0}$ & $4g_0^2/\kappa\Gamma_{\t{m}}$ & $(0.28,0.35)$ (Fig. 2) &  Inferred from $\{g_0,\kappa,\Gamma_{\t{m}}\}$.  Uncertainty taken from $\{g_0,\Gamma_{\t{m}}\}$.
\tabularnewline
\hline 
$P_\t{in}$ & & variable $(-3,+3)\%$ & Power meter at cryostat fiber output, corrected for fiber throughput loss. Uncertainty due to unkown origin of loss.
\tabularnewline
\hline 
$n_{c}$ & $\frac{4 P_\t{in}}{hc/\lambda}\frac{\kappa-\kappa_0}{\kappa}\frac{1}{1+\gamma^2/\kappa^2}$& variable $(-3,+3)\%$ & Inferred from $\{P_\t{in},\lambda,\kappa,\kappa_0,\gamma\}$. Uncertainty taken from $P_\t{in}$.
\tabularnewline
\hline
$n_{\t{tot}}$ & $\frac{S_x(\Omega)-S_x^{\t{imp}}}{2S_x^{zp}}\frac{(\Omega-\Omega_{\t{m}})^2}{(\Gamma_{\t{m}}/2)^2}$ & variable $(-6,+6)\%$ (Fig. 2)\newline $(2.2,2.7)\cdot10^4$ (Fig. 3) & Assume $n_{\t{tot}}\gg n_{\t{imp}}$.  Fit to off-resonant tail ($|\Omega-\Omega_{\t{m}}|\gg\Gamma_{\t{m}}$) of thermomechanical noise peak.  Bootstrap to $n_{\t{tot}}\approx n_{\t{th}}$ for small $P_\t{in}$.  In Fig. 2, uncertainty is taken from $n_{\t{th}}$.  In Fig. 3, additional uncertainty arises from discrepancy between two separate measurements of extraneous back-action heating (Sec. \ref{sec:error}). 
\tabularnewline
\hline 
$n_{\t{imp}}$ & $S_x^{\t{imp}}/2S_x^{\t{zp}}$& variable $(-7,+7)\%$ (Fig. 2)& Same as above.  Uncertainty taken from $\{n_{\t{th}}, \Gamma_{\t{m}}\}$.  
\tabularnewline
\hline 
$n_{\t{eff}}$ &  $\frac{S_y(\Omega_{\t{eff}})+S_x^{\t{imp}}}{2S_x^{\t{zp}}}\cdot\Gamma_{\t{eff}}$ & variable $(-12,+12)\%$ (Fig. 3) & Fit to in-loop thermomechanical noise peak under the approximation $\Gamma_{\t{eff}}\gg\Gamma_{\t{m}}$.  Bootstrap to $n_{\t{eff}}=n_{\t{tot}}\cdot\Gamma_{\t{m}}/\Gamma_{\t{eff}}$ for small $S_x^{\t{imp}}/S_y(\Omega_{\t{eff}})$.  Uncertainty taken from $\{n_{\t{tot}},\Gamma_\t{m}\}$.
\tabularnewline
\hline 
$n_{\t{eff,fb}}$ & $\frac{S_y^{\t{imp}}}{2S_y^{\t{zp}}}\cdot\Gamma_{\t{eff}}$ & variable $(-12,+12)\%$ (Fig. 3) & Same as above.
\tabularnewline
\hline 
\end{tabular}

\caption{Experimental values and their systematic uncertainties for Figs. 2-3 of the main text.}
\label{tab:Uncertainties}
\end{table*}



%

\end{document}